\documentclass[journal=jacsat,manuscript=article]{achemso}
\usepackage[position=t,singlelinecheck=off]{subfig}

\usepackage{caption}
\usepackage{color} 

\author{Srivastav Ranganathan}
\author{Eugene Shakhnovich}
\email{shakhnovich@chemistry.harvard.edu}
\affiliation[Harvard University]
{Department of Chemistry and Chemical Biology}

\title[An \textsf{achemso} demo]
{Liquid-liquid microphase separation leads to formation of membraneless organelles}

\begin{document}

% \begin{tocentry}
% 
% Some journals require a graphical entry for the Table of Contents.
% This should be laid out ``print ready'' so that the sizing of the
% Inside the \texttt{tocentry} environment, the font used is Helvetica
% 8\,pt, as required by \emph{Journal of the American Chemical
% The surrounding frame is 9\,cm by 3.5\,cm, which is the maximum
% permitted for  \emph{Journal of the American Chemical Society}
% graphical table of content entries. The box will not resize if the
% content is too big: instead it will overflow the edge of the box.
% 
% This box and the associated title will always be printed on a
% separate page at the end of the document.
% 
% \end{tocentry}

\begin{abstract}
Proteins and nucleic acids can spontaneously self-assemble into membraneless droplet-like compartments, both \textit{in vitro} and \textit{in vivo}. A key component of these droplets are multi-valent proteins that possess several adhesive domains with specific interaction partners (whose number determines total valency of the protein) separated by disordered regions. Here, using multi-scale simulations we show that such proteins self-organize into micro-phase separated droplets of various sizes as opposed to the Flory-like macro-phase separated equilibrium state of homopolymers or equilibrium physical gels.  We show that the micro-phase separated state is a dynamic  outcome of the interplay between two competing processes: a diffusion-limited encounter between proteins, and the dynamics within small clusters that results in exhaustion of available valencies whereby all specifically interacting domains find their interacting partners within smaller clusters, leading to arrested  phase separation. We first model these multi-valent chains as bead-spring polymers with multiple adhesive domains separated by semi-flexible linkers and use Langevin Dynamics (LD) to assess how key timescales depend on the molecular properties of associating polymers. Using the time-scales from LD simulations, we develop a coarse-grained kinetic model to study this phenomenon at longer times. Consistent with LD simulations, the macro-phase separated state was only observed at high concentrations and large interaction valencies. Further, in the regime where cluster sizes approach macro-phase separation, the condensed phase becomes dynamically solid-like, suggesting that it might no longer be biologically functional. Therefore, the micro-phase separated state could be a hallmark of functional droplets formed by proteins with the sticker-spacer architecture.
\end{abstract}

\section*{Significance Statement}
Membraneless organells (MO) are ubiquitous in ‘healthy’ living cells, with an altered state in disease. Their formation is likened to liquid-liquid phase separation (LLPS) between MO-forming proteins. However most models of LLPS predict complete macrophase separation while in reality MO’s are small droplets of various sizes, which are malleable  to rapid morphological changes. Here we present a microscopic multiscale theoretical study of thermodynamics and kinetics of formation of MO. We show that MO’s are long-living dynamic structures formed as a result of arrested macrophase separation. Our study provides a direct link beween the molecular  properies of MO-forming proteins and the morphology and dynamics of MO paving a path to rational design and control of MO.

\section*{Introduction}
Biomolecular phase transitions are widespread in living systems. Transitions that result in irreversible, solid-like assemblies such as amyloid fibrils are a hallmark of disease while those like cytoskeletal filaments play a functional role~\cite{Banani2017}. The third self-assembled state, in addition to the soluble and the solid-like state, is the loosely associated droplet phase held together by several weak, transient interactions~\cite{Guo2015}. The transient nature of these interactions makes these self-assemblies reversible and thereby a potential strategy for temporally regulated sub-cellular organisation. Several examples of spatiotemporally regulated, droplet-like objects within the cell have been discovered in the past few decades, composed of different types of proteins often co-localised with nucleic acids~\cite{Hyman_2014,Banani2017}. The importance of these membrane-less compartments is potentially two-fold: (i) localising biochemical reactions within the cell, and (ii) sequestering biomolecules to regulate their activity~\cite{Hyman_2014,Banani2017}. Examples of cytoplasmic membrane-less organelles include P bodies, germ granules and stress granules (SGs)~\cite{Mitrea2016}. However, aberrant granule dynamics and a transition from a liquid-like to a more solid-like state are often hallmarks of degenerative diseases.~\cite{Kim2013,RAMASWAMI2013727} Solid-like RNP aggregates have been reported as cytoplasmic inclusions ~\cite{Patel2015} and as nuclear RNP aggregates ~\cite{RAMASWAMI2013727} in degenerative diseases. Investigating the physical principles governing the formation of these high-density phases is vital to understand the subcellular organisation and the conditions leading to disease. 

Several experimental studies have highlighted the 'multi-valent' nature of the constituent proteins in membrane-less organelles ~\cite{Banani2017,Li2012,Xing2018}. In other words, these proteins carry multiple associative or 'adhesive' domains~\cite{Patel2015,Li2012}. Multivalency could be achieved by several different architectures~\cite{Banani2017}, the simplest being a linear sequence of folded domains that are connected by linker regions that are flexible and unstructured. Li et al., employed such a linear multivalent 2-component model system (SH3 and PRM domains threaded together by flexible regions) to show that liquid-like droplet formation can result from just two multivalent interacting components (repeats of the same domain)~\cite{Li2012}. Another intriguing feature of condensate proteins is the presence of intrinsically disordered regions or low complexity sequences that link folded domains together~\cite{Harmon2017}. Therefore, a combination of these motifs and different architectures could form a basis for different types of phase-separated structures within the cell. Due to the heterogeneous composition of droplets and complex set of factors dictating their growth, theoretical and computational models of varying resolutions have previously been employed to study this problem~\cite{Harmon2017,Auer,Choi2019,Ren2001}. However, much of the focus has been on identifying the self-assembled state at equilibrium. In particular, Flory theory of phase separation in polymer solution has been employed to describe self-assembly of membrane-less organelles~\cite{Brangwynne2015}. The Flory theory which is applicable to solutions of homopolymers predicts two phases – mixed and macrophase separated. However, in reality membrane-less organelles are droplets of finite size akin to microphase separated entities – in variance with simple Flory-type theoretical predictions. Despite significant experimental and computational efforts, a key question remains unanswered -- what are the physical mechanisms that govern the formation of tunable microphase separated droplets rather than complete macrophase separation? The multi-component nature of membrane-less organelles is an important feature that distinguishes them from simple homopolymers that undergo phase separation.  How do droplet size distributions depend on the intrinsic characteristics of the biopolymers?  What are the key features inherent to the architecture of these self-assembling molecules that make their assemblies tunable? Since the equilibrium states in such systems are either mixed or macrophase separated, the kinetic factors must play a crucial role in the observed microphase separation outcome~\cite{Li2012}. However, a detailed mechanistic understanding of the process of formation and growth of these droplets that leads to the microphase separated state has been lacking.

In this work, we address these questions using multi-scale coarse-grained models. In the first part of the paper, we probe the physical determinants of intracellular microphase separation using microscopic model of multivalent polymers composed of a linear chain of adhesive domains separated by semi-flexible linkers. The results of our Langevin dynamics (LD) simulations for this model shed light on the early stages of droplet growth and the mechanism of arrested phase-separation. Next, using the LD simulations as the basis to identify vital timescales for condensate growth, we explore the phenomenon at biologically relevant timescales using a phenomenological kinetic model. Broadly, we explore liquid-liquid microphase separation (LLmPS) on multiple scales – from mesoscale to macroscopic. To that end, detailed LD simulations explore microscopic mechanisms that determine key events dictating the early stages of growth which serve as the basis for developing the phenomenological kinetic model. Overall, using these two approaches, we provide a mechanistic explanation for the predominance of the microphase separated state for multi-valent heteropolymers.

\begin{figure}
\centering
\captionsetup[subfigure]{labelformat=simple}
\includegraphics[scale=0.33]{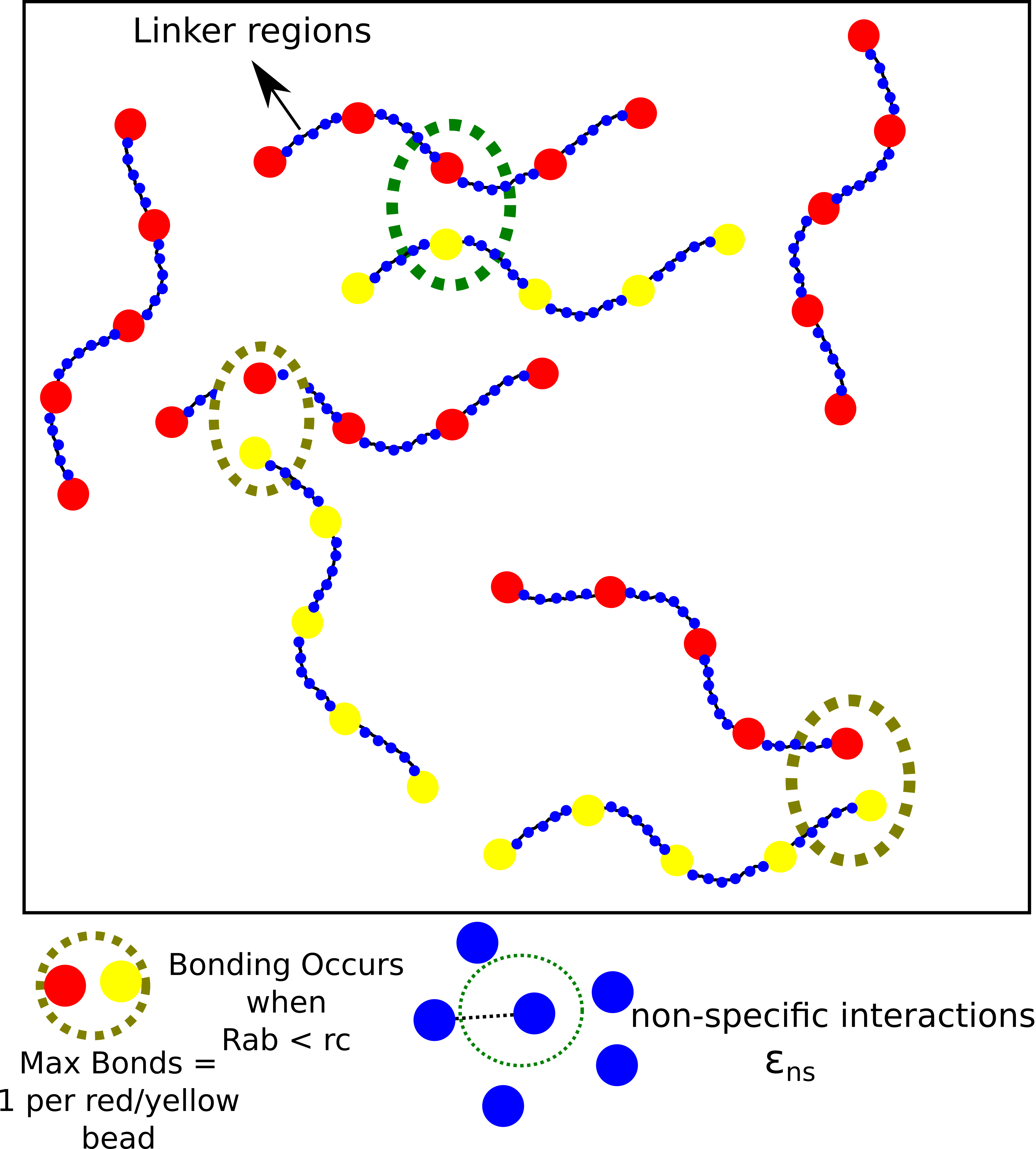}
% \hfill
% \subfloat[]{\includegraphics[scale=0.3]{Figures/growth_timescales.eps}}
\caption{Model. Schematic of the polymer model for studying phase-separation by multivalent biopolymers}
\label{fig:model_scheme}
\end{figure}

\section*{Model}
Despite the complexity of the intracellular space, experiments suggest that \textit{in vitro} LLmPS can be achieved even using simple two-component systems~\cite{Li2012}. The tractability of simpler models makes them powerful tools to investigate the role of individual factors in modulating droplet formation and growth. Here, we perform LD simulations (see Methods for detail) to understand the process of self-association between two types of polymer chains composed of specific interaction sites (red and yellow beads in Fig.1). These adhesive sites are linked together by non-specifically interacting linkers (blue beads in Fig.1). The red and yellow beads on these chains mimic complementary domains on different chains that can participate in a maximum of one specific interaction (between yellow and red beads). For our simulations, we consider 400 such semi-flexible polymer chains (200 of each type) in a cubic box with periodic boundary conditions). Each chain in the simulation box is composed of 5 specific interaction sites that are linked together by non-specific linker regions that are 35 beads long. (blue beads in Fig.1). This linker length was based on previous theoretical studies of phase-separating proteins~\cite{Harmon2017}. The specificity of the functional interactions is modelled by imposing a valency of 1 between different complementary specific bead types (red and yellow beads) via a bonding vector attached to each bead. Valency 1 means that each specific interaction site can only be involved in one such interaction. The total valency of an individual polymer chain ($\lambda$) is the number of adhesive sites that are part of a single chain. Bond formation (modelled using stochastically forming harmonic springs) occurs with a probability ($P_{form}$) if two complementary beads are within a defined interaction cutoff distance ($r_{c}$). We employ conventional Langevin dynamics simulations to study the self-assembly of the model biopolymers, wherein the size of the specific interaction sites (diameter of 20 $\AA$) is roughly four times that of the linker beads which represent individual amino acid residues (diameter of 4.2 $\AA$). This difference in sizes is to mimic a folded adhesive domain -- SH3 domain (diameter of 20 $\AA$, PDB ID:1SHG,) that is often involved in liquid-liquid phase-separation~\cite{Musacchio1992}. The folded adhesive domains are modeled at a lower resolution (one bead per domain of 60 amino acids) than linkers which are more dynamic and hence modeled with a 1 bead per amino acid resolution.

\section*{Results}
Self-assembly of multi-valent polymers with inter-chain interactions between complementary domains is a simple model system for studying phase-separation by intracellular polymers. In the current study, we employ this model to discover microscopic factors driving phase separation. In the first part of this study, we present results of self-assembly driven by irreversible (highly stable) functional interactions and identify two key timescales that define cluster growth and their size distributions. In the latter part of this paper, we build a coarse-grained kinetic model demonstrating the tunability of this phenomenon. In the following subsections, we use the term specific interactions for the finite valency interactions between functional domains and non-specific interactions to refer to the inter-linker isotropic interactions.

\begin{figure*}
\centering
\captionsetup[subfigure]{labelformat=simple}
\includegraphics[scale=0.65]{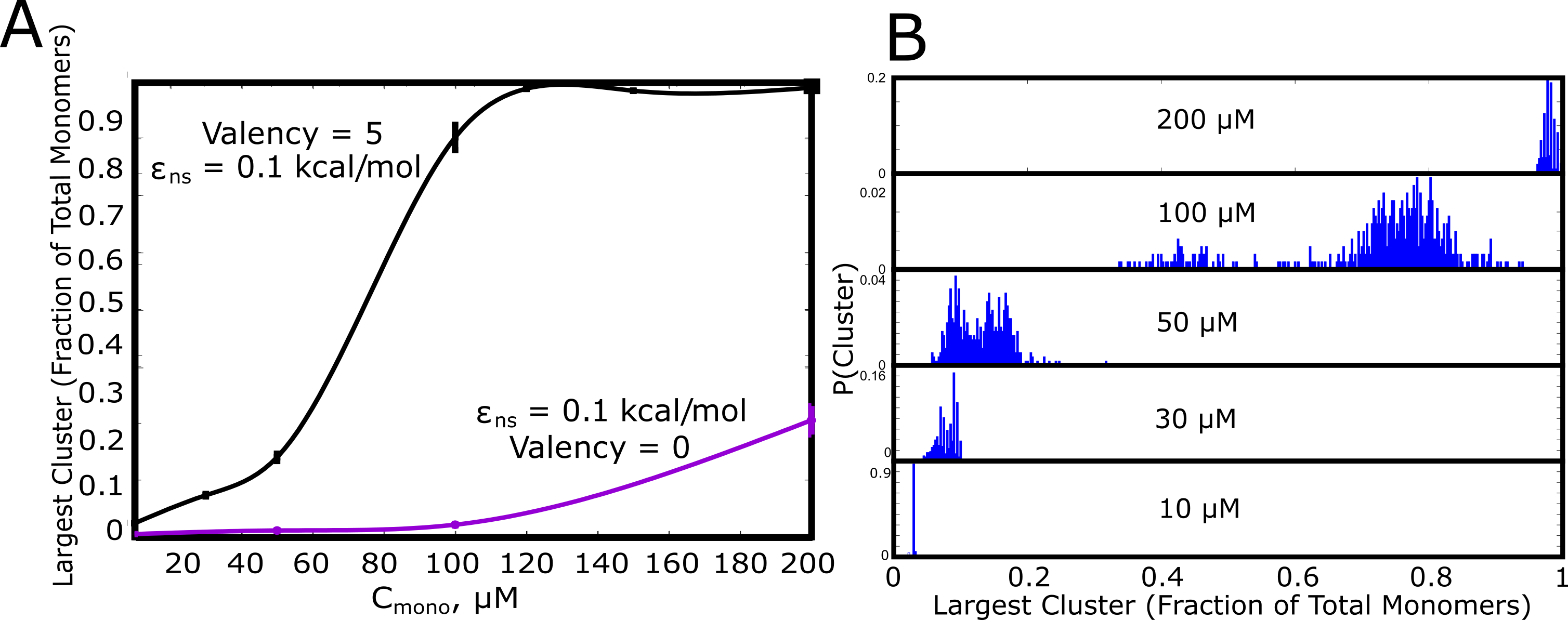}
% \hfill
% \subfloat[]{\includegraphics[scale=0.3]{Figures/growth_timescales.eps}}
\caption{Cluster Sizes in Langevin Dynamics simulations. A) The single largest cluster as a function of free monomer concentration (in $\mu$M). The largest cluster size is shown as a fraction of the total number of monomers in the simulation box. The smooth curves are plotted as a guide to the eye, using the cspline curve fitting. The vertical bars represent the standard error. B) Cluster size distributions for increasing free monomer concentrations. The linker stiffness for the self-assembling polymer chains in this plot is 2 kcal/mol while the strength of inter-linker interaction is 0.1 kcal/mol (per pair of interacting beads). The mean and distributions of the single largest cluster sizes were computed using 500 different configurations from 5 independent simulation runts of 16 $\mu$s. The simulations were performed for an instantaneous bond formation assumption ($P_{form}$ = 1).}
\label{fig:LD_clus_distri}
\end{figure*}

\subsection*{Micro-phase separation -- the most likely outcome at low concentrations and irreversible functional interactions}

Multivalent polymers with adhesive domains separated by flexible linkers can potentially self-assemble through two types of interactions, a) the finite number of adhesive contacts between the functional domains (yellow and red beads in Fig.1), and b) the non-specific, isotropic interactions between the linker regions (blue beads in Fig 1). We first performed control simulations with specific interactions turned off, where we varied free monomer concentration $C_{mono}$, from 10 to 200 µM for a weak non-specific interaction strength of $\epsilon_{ns}$ = 0.1 kcal/mol. In these control simulations, we observe no phase-separation in the whole range of $C_{mono}$ (Fig.2A, purple curve). Therefore, in this regime of $\epsilon_{ns}$ and $C_{mono}$, at the simulation timescale of 16 µs, the polymer assemblies do not reach large sizes. Not surprisingly, this result suggests that the assembly driven by non-specific interactions (linker-driven) alone is achieved only at stronger non-specific interactions and/or high free monomer concentrations.  However, intracellular microphase separation often results in the enrichment of biomolecules within the self-assembled phase, at relatively low bulk concentrations~\cite{Xing2018}. Under such conditions, specific interactions which are fewer in number become critical determinants of microphase separation. In our LD simulations, we employ polymer chains comprised of 5 univalent adhesive domains and 4 linker regions totalling 145 beads bringing total valency of 5 for each polymer. For the sake of simplicity, we begin with a situation where these bonds, once formed, do not break for the rest of the simulation. Although an idealized construct with respect to biological polymers, these simulations can aid in our understanding of cluster size distributions and arrest of cluster growth when the spontaneously assembled clusters are unable to undergo further reorganization.

In Fig.2 we show the single largest cluster size, for varying free monomer concentrations. The polymer-chains, for the data plotted in Fig.2, have flexible linkers ($\kappa$=2 kcal/mol, see Model and Methods for defintion of bending rigidity) with weak inter-linker interactions ($\epsilon_{ns}$ = 0.1 kcal/mol). As evident from Fig.2, the polymer chains can assemble into larger clusters in the presence of functional interactions (black curve, Fig.2A), as opposed to the control simulations where inter-linker interactions alone drive self-assembly (purple curve, Fig.2A). Simulations with highly stable functional interactions suggest that, for low and intermediate concentrations ($C_{mono}$ $<$ 100 uM) these polymer chains would never achieve aggregate sizes that approach the number of chains in the system (fraction of monomers in largest cluster $\rightarrow$ 1 in Fig.2A) indicating a microphase separation regime. However, an increase in concentration from 10 $\mu$M to 100$\mu$M results in macrophase separation - a system-spanning aggregate with cluster sizes approaching the total number of monomers. This is further evident from the size-distribution of the single-largest cluster computed using 500 different configurations from 5 independent simulation runs of 16 $\mu$s (Fig.2B). The largest cluster approaches the total number of monomers, only at large $C_{mono}$. However, the cellular concentration of phase-separating proteins is often in the nanomolar to the low micromolar range ~\cite{Xing2018}. Our LD simulations suggest that for irreversible specific interactions macro-phase separation can only be observed at large non-physiological concentrations.

\begin{figure*}
\centering
\captionsetup[subfigure]{labelformat=simple}
\includegraphics[scale=0.5]{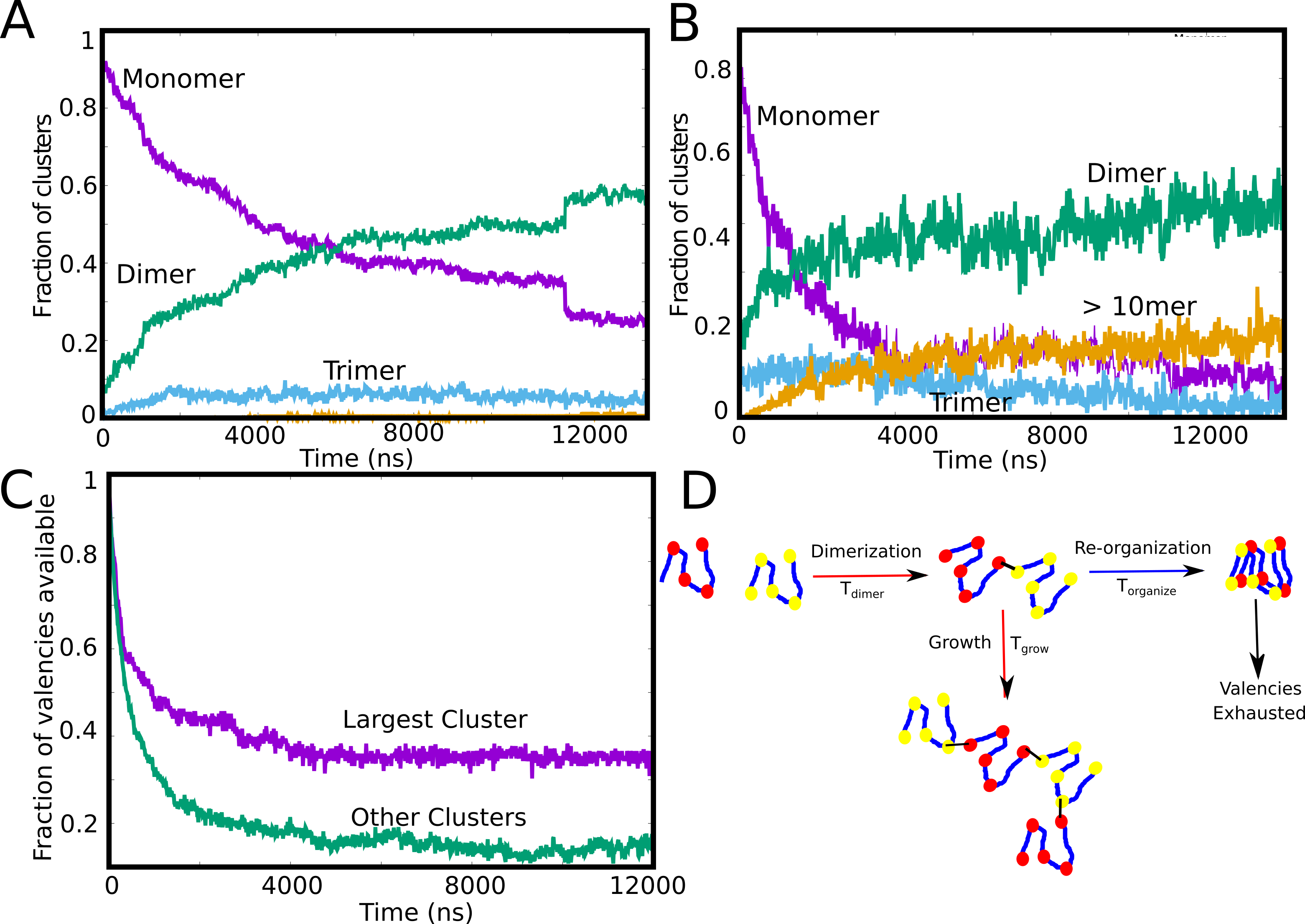}
% \hfill
% \subfloat[]{\includegraphics[scale=0.3]{Figures/growth_timescales.eps}}
\caption{Tracking cluster formation at early timescales. A) and B) show the temporal evolution of specific contacts for a free monomer concentration of 10, and 50 $\mu$M, respectively. For a low concentration of 10 $\mu$M, there is an initial decrease in the monomer population (purple curve) which is concomitant with an increase in the dimer population. A negligible fraction of the clusters is in the form of large-mers (size $>=$ 10, orange curve) at these low concentration since the available valencies for growth are consumed by the smaller aggregated species. An increase in concentration from 10$\mu$M to 50$\mu$M results in an increase in the large-mer (orange curve, 50 $\mu$M) population as the monomer fraction decreases during the simulation. A higher free monomer concentration allows the larger clusters to grow due to consumption of free monomers (with unsatisfied valencies) before they get converted into smaller clusters (dimers, trimers) with satisfied valencies. C) Time evolution of available valencies within the single largest cluster and outside the single largest cluster, for a $C_{mono}$ of 50 $\mu$M, $\epsilon_{ns}$ of 0.1 kcal/mol and linker bending rigidity of 2 kcal/mol. D) A schematic figure showing the possible mechanisms of cluster growth and arrest and the competing timescales that could punctuate the process.}
\label{fig:valency_exhaust}
\end{figure*}

\subsection*{Exhaustion of free valencies results in kinetically arrested droplets.}
The size of the largest cluster shows a dependence on concentration (Fig.2), for weak inter-linker interactions ($\epsilon_{ns}$ = 0.1 kcal/mol). While the macro-phase separated state could be a potential equilibrium outcome, for stable functional interactions, such a state is not observed in our simulations except for very high concentrations ($>=$ 100 $\mu$M in Fig.2). Therefore, identifying the timescales which are vital for cluster growth could reveal the cause of arrested macro-phase separation. In Fig.3A and B, we show the time evolution of the individual species at different monomer concentrations (10 $\mu$M and 50 $\mu$M). As seen from Fig.3A and B, for irreversible functional interactions, the monomer fraction continues to monotonically decrease during the simulations with the fraction of other competing species increasing concomitantly. However, at low concentrations (Fig.3A, 10 $\mu$M), the monomer fraction curve shows a cross-over with the dimer curve (Fig.3A, green curve) while higher order clusters do not appear at simulation tmescales.. This result suggests that the spontaneous formation of large assemblies held together by functional interactions is contingent upon two timescales. The first is the diffusion-limited timescale that governs initial dimerisation and the subsequent growth of these smaller clusters. The second, competing timescale is the one where all functional valencies get exhausted within the smaller initial clusters. At an intermediate concentration of 50 $\mu$M, (Fig.3B) we observe that the fraction of large aggregates ($>$ 10mer) increases during the early part of the simulations before free monomers are entirely consumed within dimers. In other words, the unsatisfied valencies within the monomers, dimers and trimers get utilized to form larger-sized clusters before the specific valencies get exhausted within the smaller clusters making them no longer available for further self-assembly. To further verify this observation, we track the temporal evolution of the fraction of available valencies within and outside the single largest cluster (Fig.3C). We observe that while there are unutilized valencies within the single largest cluster (Fig.3C, purple curve), the cluster does not grow further due to almost complete exhaustion of valencies within the polymer chains outside the single largest cluster (Fig.3C, green curve). A slower rate of exhaustion of free-valencies within micro-clusters by varying the specific bond formation probability, $P_{form}$, results in larger cluster sizes for smaller free monomer concentrations, suggesting that the dynamicity of bond formation can alter the cluster size distributions significantly (Supplementary Fig. S1). These results suggest that the ability to form a large, macro-cluster is limited by the exhaustion of free valencies within smaller sized clusters, thereby arresting their growth (Fig.3D). 

\begin{figure*}
\centering
\captionsetup[subfigure]{labelformat=simple}
\includegraphics[scale=0.6]{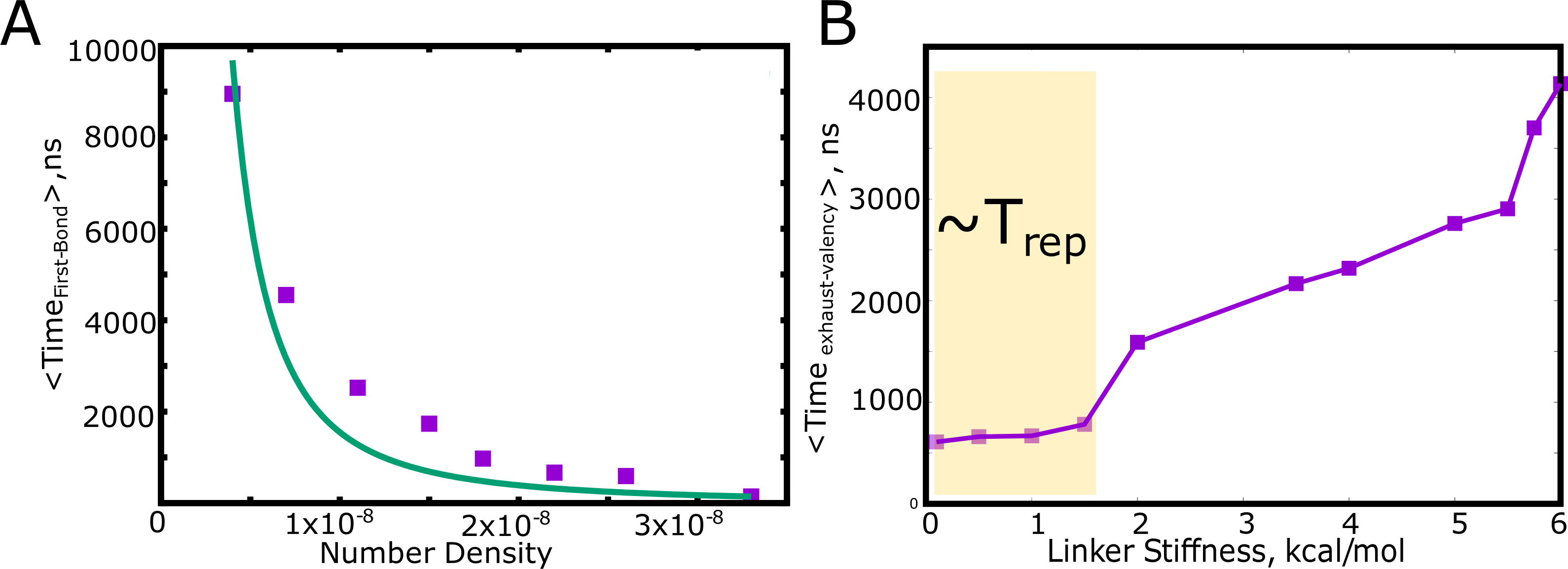}
% \hfill
% \subfloat[]{\includegraphics[scale=0.3]{Figures/growth_timescales.eps}}
\caption{The characterisation of time-scales associated with the growth of a cluster using a dimerisation model. A) The mean first passage time for the first specific interaction between a pair of polymer chains as a function of the polymer concentration (represented as the polymer number density, $N_{mono}$/$L^3$, where $N_{mono}$ and L are, the number of chains the length of the box, respectively). B) The mean first passage time for a pair of polymer chains to exhaust all the available valencies within the dimer. The shaded region in B) shows that for flexible linkers, this timescale for the exhaustion of valencies is of the order of the reptation time for an equivalent polymer, given by $T_{rep}$. $T_{rep} = \frac{L^{2}}{3*\pi*D_{bead}}$ , where L and $D_{bread}$ refer to the contour length and the diffusion coefficient of the individual monomer beads within the polymer chain~\cite{Feldman1989}.}
\label{fig:2_timescales}
\end{figure*}

\subsection*{Identifying the vital timescales determining cluster growth}
In the limit of irreversible functional interactions, the process of phase separation gets arrested due to kinetically trapped clusters which do not participate in further cluster growth due to lack of available valencies (Fig.3D). As discussed above, two critical timescales would then dictate the growth of clusters: i) the timescale for two chains to meet and form the first functional interaction, and ii) the time it takes for the polymer chains within an assembly to exhaust all valencies within the cluster before new chains join in. We explore the scaling behaviour of these two timescales using the primary unit of any self-assembly process, the dimer. Using LD simulations, we first compute the mean first passage times for two polymer chains to form their first functional interaction. Given the diffusion-limited nature, this timescale scales with the concentration of free monomers in the system (Fig.4A). This diffusion limited timescale dictates the encounter probability of the two chains. Once the first bond is formed, resulting in an 'active dimer' (one that has unsatisfied valencies), the cluster can only grow to larger sizes as long as the dimer remains active. Therefore, the time taken by the dimer to exhaust all its valencies becomes a vital second timescale. In Fig.4B, we plot the mean first passage times for a dimer to exhaust all its valencies once the first bond is formed. For flexible linkers ($\kappa$ $<$ 2 kcal/mol), this timescale (referred to as the reorganisation time, $T_{exhaust-valency}$), is roughly of the order of the reptation time of flexible polymer chains ($T_{rep}$ In Fig.4B) that are 145 beads long. The reptation time, i.e. the time taken for the polymer chain to 'scan' its entire contour length, is given by~\cite{Feldman1989},
\begin{equation}
T_{rep} = \frac{L^{2}}{3*\pi*D_{bead}},
\end{equation}
where L and $D_{bead}$ refer to the contour length and the diffusion coefficient of the individual monomer beads within the polymer chain. Also, $D_{bead}$ = $\frac{k_{B}T}{6\pi\eta a}$. For an amino acid, $D_{bead}$ is of the order of $9*10^{-10} m^{2}/s$. For a polymer of contour length $L = 145$ beads in length (each bead representing an amino acid), this timescale is of the order of 0.5$\mu$s. As seen from the right panel in Fig.4, the flexible polymer chains indeed exhaust their valencies at timescales comparable to the reptation time for an equivalent flexible polymer chain. As the linker region becomes more rigid, this timescale becomes slower, resulting in the dimer remaining `active' for much longer. A slower reorganisation time and a faster diffusion encounter time favours the cluster growing into larger sizes. Conversely, a faster reorganisation time that is of the order of the meeting timescales results in a higher likelihood of the clusters getting locked at smaller sizes. For stable functional interactions, the interplay between these two timescales would limit the size of the largest cluster. 

\begin{figure*}
\centering
\captionsetup[subfigure]{labelformat=simple}
\includegraphics[scale=0.45]{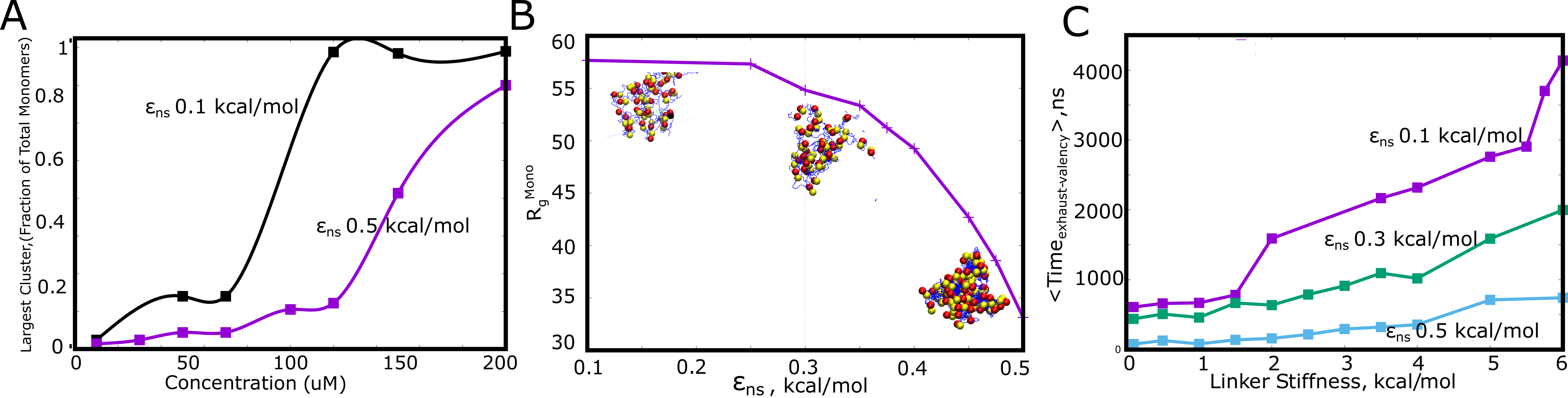}
% \hfill
% \subfloat[]{\includegraphics[scale=0.3]{Figures/growth_timescales.eps}}
\caption{Linkers as modulators of self-assembly propensity. A) The size of the largest cluster for flexible linker regions($\kappa$ = 2 kcal/mol) with varying inter-linker interaction strength (black curve, $\epsilon_{ns}$ = 0.1 kcal/mol and purple curve, $\epsilon_{ns}$ = 0.5 kcal/mol). Sticky inter-linker interactions result in smaller cluster sizes, also corroborated by the collapsed nature of these aggregates in B). Mean Radius of gyration for the individual-polymer chains ($Rg^{Mono}$) within a self-assembled cluster as a function of increased inter-linker interactions. C) The mean reorganisation times ($T_{exhaust-valency}$,  as a function of linker-stiffness, for different values of inter-linker interaction strengths.}
\label{fig:linker_modulate}
\end{figure*}

\subsection*{Linker regions as modulators of self-assembly propensity}
The spatial separation of specifically interacting domains (with finite valencies) and the non-specific linker regions is an architecture that could be highly amenable to being tuned for phase-separation propensities. In this context, we further extend the findings from our LD simulations to probe how the microscopic properties of the linker region could modulate the extent of phase-separation. In Fig.5, we demonstrate how the linker properties could be useful modulators of cluster sizes, without any alteration of the nature of the specific functional interactions. Further, to model an unstructured linker, we consider a scenario where the linkers participate in inter-linker interactions alone. Strong inter-linker interactions increased occupied valencies (Supplementary Fig.S2), for all the concentrations under study, while resulting in much smaller equilibrium cluster sizes (Fig.5A) during the timescales accessed by our simulations. Further, the mean radius of gyration ($Rg^{Mono}$) for the monomers within these clusters shows a transition with an increase in inter-linker interaction strength ($\epsilon_{ns}$), with a sharp decrease in $Rg^{Mono}$ for an increase in $\epsilon_{ns}$ from 0.3 kcal/mol to 0.5 kcal/mol indicating an onset of linker-driven coil-globule transitions for the polymers~\cite{Lifshitz1978}.

The assemblies that ensue at these strong values of $\epsilon_{ns}$ are more compact and condensed akin to homopolymer globules~\cite{Lifshitz1978}. Intramolecular compaction of polymers due to nonspecific inter-linker interactions brings specific domains closer in space leading to a higher likelihood of the exhaustion of specific interaction valencies within small assemblies. In Fig.5C, we show how the inter-linker interaction strength can influence the time it takes to exhaust  specific interaction valencies within a dimeric cluster ($T_{exhaust-valency}$). For weaker inter-linker attraction ($\epsilon_{ns}$ $<$ 0.5 kcal/mol), the initial polymer assemblies are less compact (Fig.5B) and thereby exhaust valencies within a cluster at a much slower rate (higher $T_{exhaust-valency}$ for dimers with $\epsilon_{ns}$ $<$ 0.5 kcal/mol in Fig.5C). An increase in inter-linker interaction propensity results in faster re-organisation times for these polymers. The polymers with $\epsilon_{ns}$ = 0.5 kcal/mol exhaust their valencies almost an order of magnitude faster than their 0.1 kcal/mol counterparts (Fig.5C). Upon exhaustion of these specific interaction valencies, these clusters can only grow via inter-linker interactions, a phenomenon that could be less dynamic (Supplementary Fig S3) and tunable than the functional interaction driven cluster growth. It must, however, be noted that the observation of further coalescence of clusters formed by sticky inter-linker interactions was limited by the timescales accessible to the LD simulations. Any alteration to the ‘stickiness’ of the linker can shift the mechanism of assembly, and thereby result in altered kinetics of cluster growth by modulating the $T_{exhaust-valency}$. Our dimerisation simulations show that a second mechanism of slowing down the $T_{exhaust-valency}$ timescale is by altering the flexibility of the linker region (increasing linker stiffness in Fig.5C). Primarily the stiffer linkers model spacer regions that prefer more `open' configurations (because of say, a high local density charges). This is corroborated by a shift in the cluster size distribution towards larger sizes, for the polymer chains with less flexible linkers (Supplementary Fig.S4). The linker region can thereby serve as a modulator of microphase-separation propensity.
Further, we probed how the linker region influences the density profiles of these clusters. In Supplementary Fig.S4, we show the probability distributions for cluster densities normalized by the bulk densities, as a measure of the degree of enrichment within the self-assembled state. For weak inter-linker interactions, we find a 10-fold enrichment of monomers within the assemblies (Supplementary Fig.S4, purple curve). On the other hand, a ‘sticky’ linker results in an almost 100-fold enrichment within the collapsed globule-like clusters (Fig.5B and Supplementary Fig.S5). These values are consistent with experimental findings of a $\approx$10-100 fold enrichment of biomolecules within droplets~\cite{Lifshitz1978,Li2012,Xing2018,Nott2015,Burke2015}. Therefore, a variation in the intrinsic properties of the linker, with no modifications to the functional region can be used as a handle to tune the degree of enrichment within the condensed phase. This lends modularity and functionality to these condensates, with the linker regions controlling the degree of enrichment within a condensate. 

\begin{figure*}
\centering
\captionsetup[subfigure]{labelformat=simple}
\includegraphics[scale=0.55]{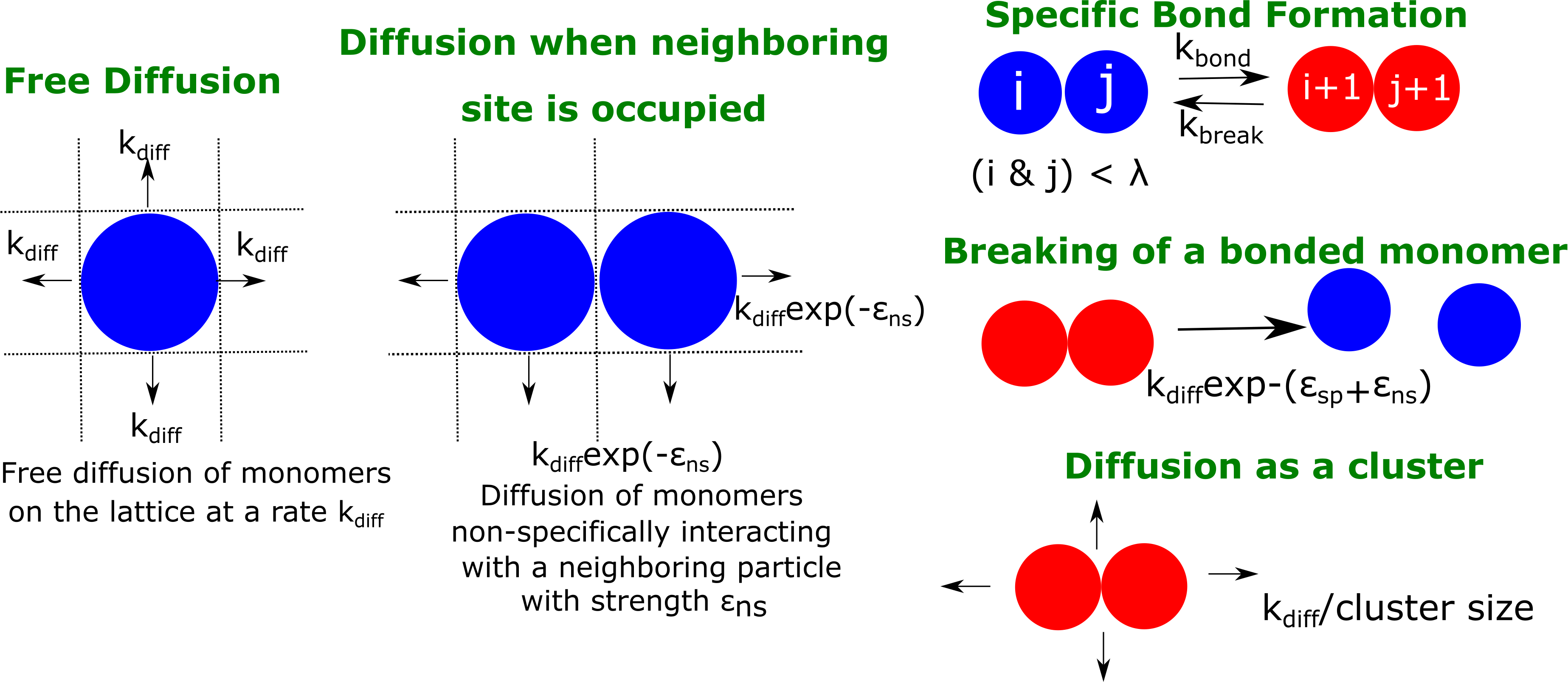}
% \hfill
% \subfloat[]{\includegraphics[scale=0.3]{Figures/growth_timescales.eps}}
\caption{A schematic figure detailing the different rates in our phenomenological kinetic model simulated using the Gillespie algorithm. The particles on the lattice can diffuse freely (when there are no neighbouring particles) with a rate $k_{diff}$. In the presence of a neighbouring particle, a non-specifically interacting monomer can diffuse away with a rate $k_{diff}$ $\epsilon_{ns}$. Neighbouring particles can also form specific interactions (with fixed valency $\lambda$) at a rate $k_{bond}$ or break an existing interaction with a rate $k_{break}$. Clusters could diffuse at a rate that is scaled by their sizes. $\epsilon_{ns}$ and $\epsilon_{sp}$ refer to the strength of non-specific and specific interactions, respectively.}
\label{fig:kmc_scheme}
\end{figure*}

\subsection*{Phenomenological kinetic simulations predict microphase separation at biologically relevant timescales} 
The LD simulations help us identify the initial events that mark phase-separation by multi-valent polymer chains assembling via finite-valency, specific interactions. However, the model is limited by its ability to access longer, biologically relevant timescales at which droplets typically form and grow in living cells. Also, while the assumption of non-transient interactions is a useful simplification to identify conditions arresting cluster growth, bio-molecular interactions are often transient in nature. Hence, to probe whether micro or macrophase separation becomes kinetically favored for reversible functional interactions, we employ a coarse-grained approach wherein the whole polymer chain from the bead-spring model (with a fixed valency) is represented as an individual particle on a 2D-lattice. In our simulations, the lattice is populated by ‘N’ such multi-valent particles (at varying densities) that diffuse freely, at a rate $k_{diff}$, and particle collisions per unit time is proportional to $k_{diff}$*$\phi$. Here, $\phi$ is the bulk number density of particles on the lattice. In our MC study, we vary the bulk density of the lattice in a range of 0.01 to 0.1 (see Methods section for rationale). When two such particles occupy neighbouring sites on the lattice, they interact non-specifically with an interaction strength of $\epsilon_{ns}$, a parameter that is analogous to the inter-linker interactions in our LD simulations. Additionally, two neighbouring particles with unfulfilled valencies can form a specific bond (with finite valencies per lattice particle) with a rate of $k_{bond}$. The valency per particle ($\lambda$) here can be utilized to form bonds with one to four potential neighbors, with each pair being part of one, or more than one specific interactions between themselves. However, unlike the irreversible specific interactions in our LD simulations, the specific bonds in the lattice model can break with a rate $k_{break}$ = $k_{bond}$*exp(-$\epsilon_{sp}$), where $\epsilon_{sp}$ is the strength of each specific bond. Additionally, clusters can diffuse with a scaled diffusion rate that is inversely proportional to the cluster size. It must be noted that the timescale for the first bond formation in the LD simulations is an outcome of two phenomenological rates in this model, $k_{diff}$ and $k_{bond}$. The second timescale, $T_{exhaust-valency}$, is a timescale that depends on the $k_{bond}$ and $k_{break}$ parameters in this model. The details of the simulation technique and the various rate processes can be found in the Methods section and described schematically in Fig.6. 

\begin{figure*}
\centering
\captionsetup[subfigure]{labelformat=simple}
\includegraphics[scale=0.7]{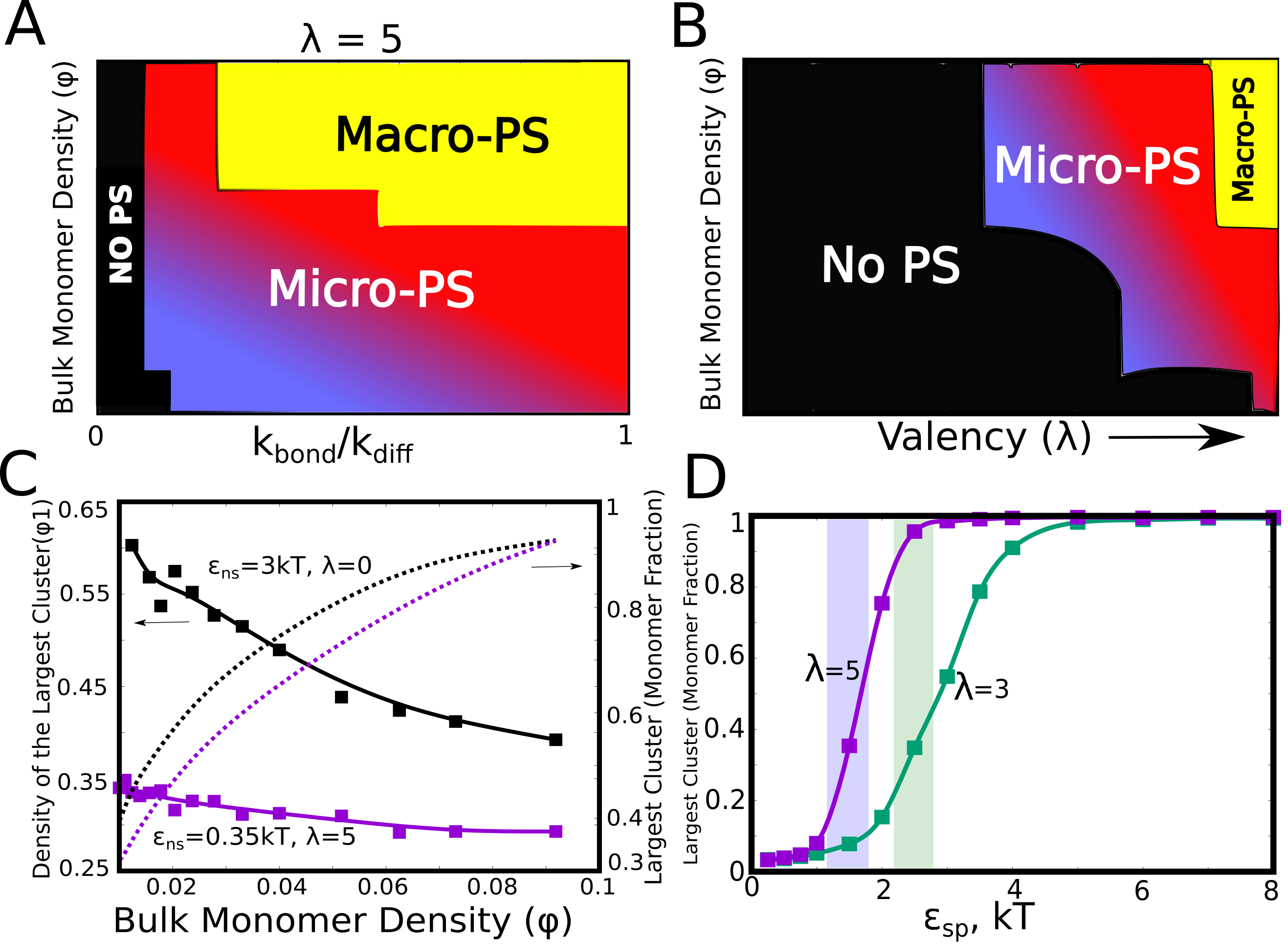}
% \hfill
% \subfloat[]{\includegraphics[scale=0.3]{Figures/growth_timescales.eps}}
\caption{Kinetic Monte Carlo Simulations. A.) Phase diagram highlighting the different phases (microphase (Micro-PS) or macrophase separated (Macro-PS), and the non-phase separated state (No PS) ) encountered upon varying the ratio of $k_{bond}$/$k_{diff}$, and the bulk density of monomers ($\phi$) within the box. The assembling particles have a valency ($\lambda$) of 5 in these simulations. B.) Phase diagram highlighting the different phases encountered upon varying valency and bulk density as the phase parameters. The macro-phase separated state is only encountered for larger valency particles at higher densities. This phase diagram was computed for $k_{bond}$/$k_{diff}$ = 1 and $\epsilon_{ns}$ of 0.35 kT. C.) The difference in the densities of the largest cluster, for clusters that are driven by specific (solid purple curve) and non-specific (solid black curve) interactions. The non-specific interaction driven clusters are denser than their specific-interaction driven counterparts. The cluster sizes, however, for both these cases are comparable (dashed curves). $k_{bond}$/$k_{diff}$ = 1, and $\phi$=0.04, for this plot. $\epsilon_{ns}$ was set to 0.35 kT. D) The fraction of monomers in the largest cluster as a function of epsilon, for $k_{diff}$= $k_{bond}$ and $\phi$=0.04. The single largest cluster sizes in all sub-panels of this Figure was computed for a simulation timescale of 2 hrs (with the fundamental timsecale of diffusion being set to $k_{diff}$= 1 s-1).}
\label{fig:kmc_pd}
\end{figure*}

Using kinetic Monte Carlo simulations~\cite{Gillespie1977}, we explore the cluster formation (at times reaching a physiologically relevant scale of hours) by varying parameters such as a) bulk density of particles on the lattice ($\phi$), (2) rate of bond formation ($k_{bond}$), (c) valency per interacting particle ($\lambda$), and (d) the strength of specific interactions ($\epsilon_{sp}$). With the assumption that, for low concentrations, the rate of free diffusion $k_{diff}$ is the fundamental timescale limiting cluster growth, we first explore the relationship between the rate of specific bond formation ($k_{bond}$), and $k_{diff}$ (Fig.7A and Supplementary Fig.S6). It must be noted that the LD simulations employed the assumption that bond formation, upon the two functional domains coming in contact, is an instantaneous event. Here, we show that for values of $k_{bond}$/$k_{diff}$ $\rightarrow$ 0, there is no phase-separation. As the bond formation rate approaches that of free diffusion – correponding to instant bond formation in LD simulations --, we encountered phase-separated states in our simulations. However, the system largely favors the micro-phase separated state (bluish-red regions in the phase diagram) for low and intermediate densities. The macro-phase separated state is only observed for higher densities. The macro-phase separated regime in this phase diagram is, however, absent at low valency $\lambda$ of 3 (Supplementary Fig.S6). Further, this phase diagram (Fig.7A) also establishes that for the value of non-specific interaction strength ($\epsilon_{ns}$ = 0.35 kT, see methods section for rationale) used here the cluster formation is driven by the finite-valency specific interactions (absence of phase separation for $k_{bond}$/$k_{diff}$ $\rightarrow$ 0). For comparison, in the Supplementary Fig.S7, we present the mean cluster sizes for assemblies that are stabilized by non-specific interactions only ($\lambda$=0). The $\epsilon_{ns}$-$\phi$ phase diagram shows that non-specific interaction-driven cluster formation occurs at only high values of $\epsilon_{ns}$ (Supplementary Fig.S7B). Therefore, microphase separation is contingent on the bonding rate being of the same order as the free-diffusion rate ($k_{bond}$ $\rightarrow$ $k_{diff}$) establishing the validity of the instantaneous bonding assumption in the LD simulations. It is the ratio of $k_{bond}$/$k_{diff}$, and not the absolute magnitudes that is a vital parameter for these simulations. Hence, in all the kinetic Monte carlo (KMC) simulations we set the value of $k_{diff}$ to 1 $s^{-1}$ and vary the ratio $k_{bond}$/$k_{diff}$ to tune phase separation. It must be noted that, unless mentioned otherwise, the results from the kMC simulations presented here are for a weak non-specific interaction strength of 0.35kT. All simulations were performed for a timescale of 2 hours (actual time). As proof of convergence of these simulations, we compare results at the end of 2 hrs to those at longer simulation timescale and show that there is negligible difference in cluster sizes (Supplementary Fig.S8). 
Further, we systematically probed the effect of specific interaction valency on the extent of phase separation. Fig.7B shows phase diagram with $\lambda$ and $\phi$ as the phase parameters. For smaller $\lambda$ and low $\phi$, the single largest cluster sizes do not approach the macrophase separated limit (blue and black regions in Fig.7B). This suggests that the critical densities for macrophase separation for lower valency particles would be extremely large, making it an improbable scenario at low \textit{in vivo} concentrations. Detailed versions of these phase diagrams can be found in Supplementary Fig.S5. This phase-diagram is consistent with \textit{in vitro} experiments involving SH3 and PRM chains with varying valencies, with higher valency molecules displaying a lower critical concentration for phase separation~\cite{Li2012}. Another interesting feature that differentiates the clusters stabilized by specific interactions are relatively lower densities of these clusters, as opposed to their counterparts stabilized by non-specific interactions (Fig.7C). While the assemblies formed by specific interactions ($\lambda$=5, $\epsilon_{ns}$ = 0.35 kT; purple curve in Fig.7C) are of comparable sizes to those driven by non-specific interactions, their densities are almost 2-fold lower. This is consistent with the spatial density profiles for strong and weakly interacting linkers in our LD simulations (Supplementary Fig.S4).  In addition to the number of specific interactions, a vital parameter that would determine the droplet sizes is the strength of these specific interactions ($\epsilon_{sp}$). As evident from Fig.7D, for specific interactions that are extremely weak ($\epsilon_{sp}$ $<$ 2kT in Fig.7D), there is no significant phase separation. Strikingly, this critical interaction strength (when the largest cluster $<$ 50$\%$  of available monomers) is lower for higher valency particles ($\lambda$ = 5 curve in Fig.7D). Interestingly, the SH3-PRM interaction strength is reported to be of the order of 2 kT~\cite{Li2012,Harmon2017,Harmon2017}. A more detailed $\epsilon_{sp}$-$k_{bond}$ phase diagram can be found in Supplementary Fig.S9. Overall, our systematic study of cluster sizes suggests that the propensity to phase separate at biologically relevant timescales could be tuned via different parameters offering the cell several handles to modulate sizes and morphologies of droplets. Interestingly, a macro-phase separated state exists only in a very narrow window of parameters in the limit of weak non-specific interactions, unlike aggregation processes where isotropic interactions always result in macrophase separation at longer timescales.

\begin{figure*}
\centering
\captionsetup[subfigure]{labelformat=simple}
\includegraphics[scale=0.6]{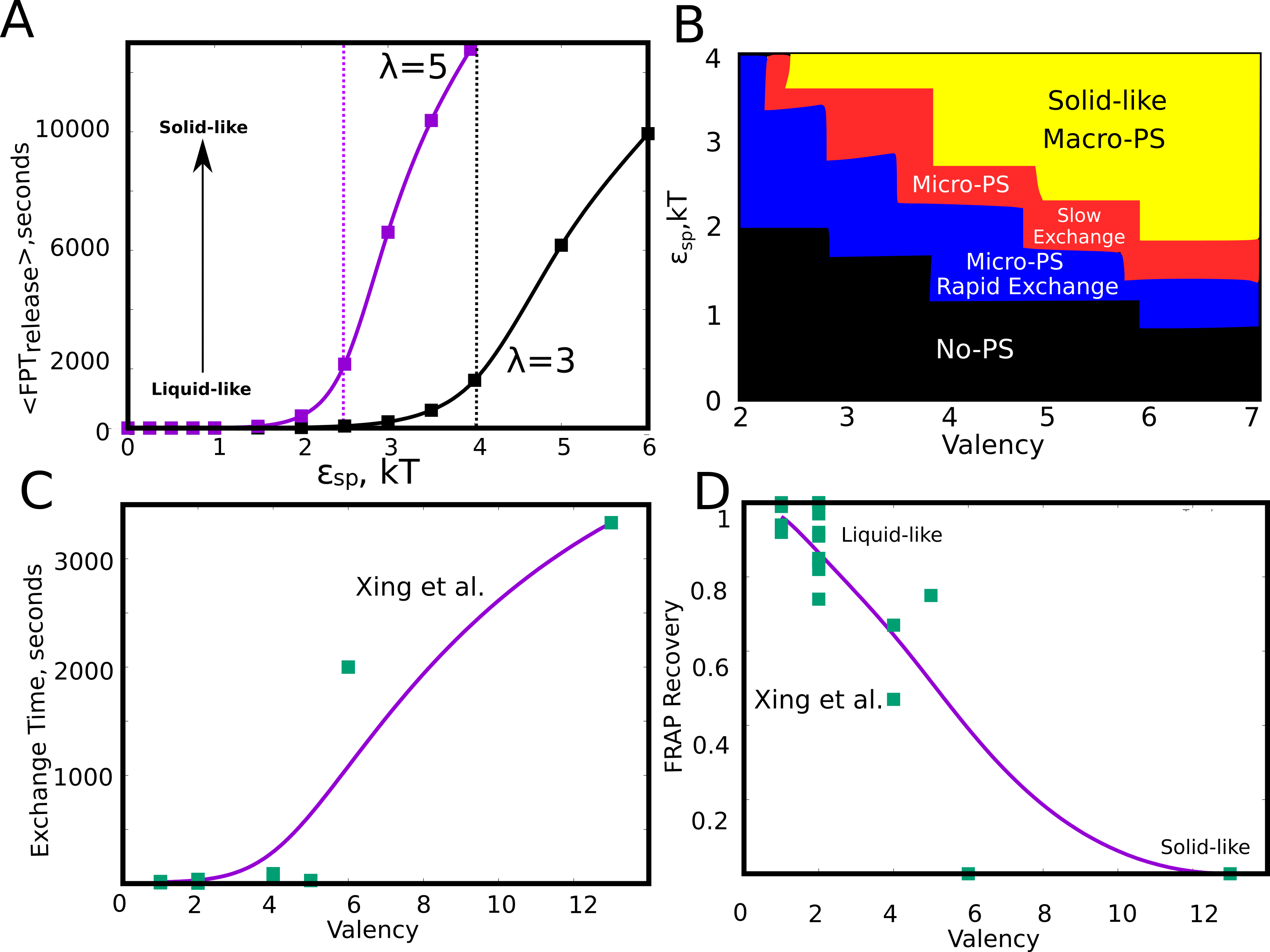}
% \hfill
% \subfloat[]{\includegraphics[scale=0.3]{Figures/growth_timescales.eps}}
\caption{Effect of model parameters on the exchange times between monomers and the aggregates. The parameter values used in panels A and B are $\phi$=0.04, $k_{bond}$/$k_{diff}$ = 1. A) Mean first passage time for the monomers to go from the buried state (with 4 neighbours) to the free state (with no neighbours) in response to varying values of specific interaction strength. The two curves show the trends for species with different interaction valencies. The dashed solid lines refer to value of $\epsilon_{sp}$ beyond which the largest cluster consumes 50$\%$ or more of available monomers. B) The state of the system, for variation in $\epsilon_{sp}$ and $\lambda$ suggests that the system displays remarkable malleability in dynamicity and size distributions. Weak $\epsilon_{sp}$ and a low $\lambda$ results in no phase separation (shaded black). For higher values of both these parameters, the system can access the macrophase separated state (shaded yellow), however with a dramatic slowdown in exchange times. For an intermediate range, the system shows microphase separation, with either slow (shaded red) or fast-exchange (shaded blue) dynamics. C) The experimentally determined molecular exchange times for molecules of varying interaction valencies.~\cite{Xing2018}. D) The extent of recovery after a photobeaching experiment, for interacting species with varying valencies. The data in panel C) and D) has been obtained from a study by Xing et al~\cite{Xing2018}. The solid curves in C and D are added to guide the eye}
\label{fig:exchange_times}
\end{figure*}

\subsection*{Tunability of exchange times in microphase-separation}
In the KMC simulations so far, we discuss the manner in which different phase parameters could shape droplet size distributions. However, the functionality of a condensate hinges not only on the ability of biomolecules to assemble into larger clusters but also to exchange with the surrounding medium at biologically relevant timescales. These exchange timescales are also a measure of the material properties of the droplets themselves~\cite{Guo2015,Xing2018,Feric2016}. Therefore a systematic understanding of the dependence of molecular exchange times on intrinsic and extrinsic parameters is crucial to get a grasp of the tunability of intracellular self-organisation driven by finite-valency specific interactions. In this context, we probed the extent to which the exchange times could be tuned by modulating the intrinsic features of the self-assembling units. Here, we define monomer exchange times as the mean first passage time for a monomer to go from having four neighbours to being completely free. To compute first passage times, we kept track of exchange events from across 300 simulation trajectories of 10 hours each. Our simulations suggest that, for a given valency, a slight increase in interaction strength ($\epsilon_{sp}$)  within a narrow window could result in a dramatic increase in phase-separation (Fig.7D). This raises an interesting question—is there an optimal range for these phase parameters that promotes phase separation as well as maintains the dynamicity of the clusters? In this context, we first computed the mean first passage times for monomer exchange upon systematic variation of $\epsilon_{sp}$ (Fig.8A). Interestingly, as with cluster sizes, a slight increase in $\epsilon_{sp}$ results in a dramatic increase in monomer exchange times. For particles with $\lambda$=5, a slight increase in $\epsilon_{sp}$ from 2 kT to 2.5 kT, there is a four-fold slowdown in exchange times indicating dramatic malleability of dynamicity of these assemblies. This shift from the fast to slow exchange dynamics is less abrupt in case of particles with $\lambda$=3 suggesting that an interplay between $\lambda$ and $\epsilon_{sp}$ could tune the droplets for desired exchange properties. We further varied these two parameters ($\lambda$ and $\epsilon_{sp}$) systematically to probe their effect on cluster sizes (Supplementary Figure.S10A) and molecular exchange times (Supplementary Figure.S10B). As expected, for weak $\epsilon_{sp}$ and a low $\lambda$ there is no phase separation (black region in Fig.8B). For an intermediate regime in this phase-space, the system is predominantly in a micro-phase separated state, with either slow (red region in Fig.8B) or fast (blue region in Fig.8B) molecular-exchange times. However, macrophase separation is only observed for really large $\lambda$ and $\epsilon_{sp}$, with a dramatic slowdown in exchange times (shaded yellow region in Fig.8B and Supplementary Figure.S10B ), suggesting that these assemblies might be biologically non-functional. Valency is, therefore, a key determinant of how frequently a molecule gets exchanged between the droplet and the free medium. Similar observations have been made experimentally by Xing et al. (Fig 8C and D) with regards to several condensate proteins featuring different valencies, with low valency species showing exchange times that are orders of magnitude faster than the higher valency ones~\cite{Xing2018}. Given that functional droplets are tuned for liquid-like behaviour, microphase separation is, therefore, the most likely outcome for dynamically exchanging droplets. 

\section*{Discussion}

\subsection*{Micro-phase separation: a potential signature of multivalent heteropolymers}
Membrane-less organelles are heterogeneous pools of biomolecules which localize a high density of proteins and nucleic acids~\cite{Banani2017,Feric2016,Boeynaems2019}. An interesting feature of several complex monomers~\cite{Mitrea2016,Li2012,Harmon2017} that constitute these droplets is 'multivalency' or multiple repeats of adhesive domains~\cite{Banani2017,Patel2015,Li2012}. These adhesive domains can bind to complementary domains on other chains, thereby facilitating phase separation. In this study, we model this phenomenon as that of self-associative polymers that possess folded domains (represented as idealized spheres in ) separated by flexible linker regions. Recent computational studies have employed similar models to characterize the equilibrium state of these polymer systems, notably the coarse-grained simulations by Harmon et al.~\cite{Harmon2017} and Choi et al.~\cite{Choi2019}. These studies employ the 'sticker and spacer' model to understand the phase behaviour of these linear multivalent polymers~\cite{Harmon2017}, largely focusing on the nature of the phase-separated state at equilibrium. Using lattice-polymer Monte carlo simulations, they establish the role of intrinsically disordered linker regions as molecular determinants that dictate the equilibrium state of a system of associative polymers that interact via non-covalent interactions~\cite{Harmon2017}. Crucially, these works focus on the cross-linked gel-like nature of the macrophase-separated state of these polymers. However, equilibrium lattice simulations and classical Flory-like theories do not address the dynamics of cluster growth resulting in the observed peculiar micro-phase separated nature~\cite{Li2012,Feric2016} of multiple biomolecular condensates, as opposed to a single macro-phase separated state~\cite{Brangwynne2015,Flory1942}. This raises a number of interesting questions – is microphase separation a signature of multi-valent protein assemblies? How does the physics of associative polymers multivalent proteins differ from those of simple, homopolymer chains? In this work, we provide a kinetic analysis of these questions using a combination of coarse-grained LD simulations to probe the early stages of cluster growth, and a reaction-diffusion model to probe the problem at longer timescales. 

\subsection*{Exhaustion of specific interaction valencies -- a barrier to macro-phase separation}
First, we studied self-assembly by multivalent polymers whose adhesive domains interact via stable, `non-transient' interactions to understand the early events in the growth process. We observed that, except for extremely high concentrations where there is a system spanning network, the most feasible scenario at smaller concentrations is that of micro-droplets with a concentration-dependent distribution of droplet sizes. The temporal evolution of aggregated species suggests that at lower concentrations, the available interaction valencies get consumed within smaller-sized assemblies, making them inert for further cluster growth. Our results suggest that two critical timescales decide whether a cluster continues to grow further -- a) concentration-dependent timescale of two chains encountering each other and forming the initial functional interaction, and, b) the exhaustion of valencies within a small cluster, a timescale dependent on intrinsic features of the polymer (Fig.4). Crucially, these two timescales are sensitive to subtle modifications in the self-assembling polymer chain (Fig.5). Therefore, by tuning these two time-scales, the cell can modulate the degree of phase separation. Modifications to the linker can also result in altered densities of the self-assembled state, with a 10-100 fold enrichment in molecular concentrations within the droplets (Supplementary Fig.S5), consistent with the experimentally reported degrees of enrichment within condensates ~\cite{Mitrea2016,Li2012,Nott2015}. Overall, the simulations involving non-transient interactions help us establish an understanding of the essential physical mechanisms determining microphase separation in membrane-less organelles, with the finite nature of the specific interactions driving the peculiar phenomenon. Although these findings were for an ‘artificial’ assumption of extremely stable functional interactions, the highly cross-linked nature of the phase-separated state~\cite{Harmon2017} might result in these early micro-phase droplets not being able to transition to the macro-phase separated state even for reversible, transient interactions.

\subsection*{Bridging the gap between the early and biologically relevant timescales}
To reach the time scales relevant to biology we employed a coarse-grained kinetic model where each polymer (from the LD simulations) is represented as a diffusing reacting centre on a 2D lattice which can interact either non-specifically (mimicking inter-linker interactions) or specifically with neighbouring centres. The difference between the two types of interactions is that the number of specific interactions that each centre can make is limited by its valency. In an extension of the LD model, specific interactions are stable yet reversible and can form and break with rates dictated by detailed balance. Consistent with our LD simulations, our kinetic Monte-Carlo simulations of the phenomenological model reveal that for timescales relevant to biology, macro phase-separation is a phenomenon that is observed in a very narrow regime of parameters in the kinetic phase-space (Fig.7). Further, the phenomenon of exhausted adhesive valencies is even more prominent for species with lower valency (fewer adhesive domains in the prototypical polymer), as evident from much smaller sizes of the largest cluster after an hour-long (actual time) simulation run. The lattice-diffusion model, despite its minimal nature, captures the well-known relationship between molecular valency and critical concentration~\cite{Li2012} for phase separation (Fig.7B). An interplay between the valency of the generalized polymer and the strength of interactions can also alter the exchange times of molecules with the bulk medium dramatically. Interestingly, a slight shift in either these valencies or interaction strengths could result in a change in exchange rates by orders of magnitude. Further, for regions of the parameter space ($\epsilon_{sp}$ and $\lambda$) that favor macro-phase separation we observe a dramatic slow-down in molecular exchange times. In other words, for parameters that result in a fast-exchanging condensed state, micro-phase separation is the most favored outcome (Fig.8B and Supplementary Fig.S10). Such a discontinuity makes these systems extremely sensitive to mutations that might cause a shift in dynamics and eventual loss of function of these droplets. Also, such shifts could also make these systems extremely responsive to non-equilibrium processes such as RNA-processing, side-chain modifications (acetylation, methylation) that are often attributed to modulating condensate dynamics~\cite{Hofweber2019,Wang2014,Wang2018}. The differential exchange times in response to variation of interaction parameters in our model lends further support to the scaffold-client model~\cite{Xing2018}. The scaffolds that are slower exchange species with higher valencies acting could, therefore, recruit faster exchange clients with lower valencies. The valencies and strength of interactions could have thus evolved to achieve exchange times that ensure the functionality of spatial segregation via liquid-liquid phase separation.
A multi-domain architecture such as the sticker-spacer architecture allows for separation of two functions with the folded domains (conserved) performing functional role while the spacer regions being modified over time to tune the propensity to phase separate and also the material nature of the condensate. Overall, our multi-scale study shows that the block co-polymer like organization of these multi-valent proteins, with finite specific interactions driving phase separation could manifest itself in the micro-phase separated droplet state. A switch in the driving force for self-assembly, from the specific to non-specific interactions via sticky linkers, could not only alter the kinetics of assembly but also have implications in disorders associated with aberrant phase separation.

% \section*{Methods}
\section{Methods}
\subsection*{Langevin Dynamics simulations}
\subsubsection*{Force Field.} The polymer chains in the box are modelled using the following interactions. Adjacent beads on the polymer chain are connected via harmonic springs through the potential function, 

\begin{equation}
E_{stretching} = k_{s}\sum\limits_{i=1}^{M-1}(|\vec{r_{i}}-\vec{r_{i+1}}|-r_{0})^2,
\label{stretch}
\end{equation}

$\vec{r_i}$ and $\vec{r_{i+1}}$ refer to the adjacent $i^{th}$ and $(i+1)^{th}$ bead positions, respectively. Here, $r_{0}$ is the equilibrium bond length and $k_{s}$ represents the spring constant. To model semi-flexibility of the polymer chain, any two neighbouring bonds within the linker regions of the polymer chains interact via a simple cosine bending potential 

 \begin{equation}
E_{bending} = \kappa\sum\limits_{i=1}^{M-2}(1-cos~\theta_{i}),
\label{bending}
\end{equation}

where $\theta_{i}$ describes the angle between  $i^{th}$ and $(i+1)^{th}$ bond while $\kappa$ is the energetic cost for bending. The non-bonded isotropic interactions between linker beads and linker-functional interactions were modelled using the Lennard-Jones (LJ) potential, 

\begin{equation}
E_{nb} = 4\epsilon\sum\limits_{i<j}\left[\left(\frac{\sigma}{|\vec{r_{i}}-\vec{r_{j}}|}\right)^{12} - \left(\frac{\sigma}{|\vec{r_{i}}-\vec{r_{j}}|} \right)^6 \right], 
\end{equation}

for all $|\vec{r_{i}}-\vec{r_{j}}|$ $<$ $r_{c}$, where $r_{c}$ to the cutoff distance beyond which the non-bonded potentials are neglected. The LJ potentials were truncated at a distance cutoff of 2.5$\sigma$. The strength of the attractive component of this potential, ε, was varied to achieve varying degrees of inter-linker interactions in our simulations.  signifies the strength of the attractive interaction and has units of energy. In our simulations, we vary the bending energy parameter ($\kappa$) to model rigid or flexible linkers. The `stickiness' of the linker regions and their ability to promote inter-molecular self-assembly is modulated using an effective interaction parameter ($\epsilon$). Non-specific, isotropic interactions between linker beads (interaction strengths ) are also used as a model parameter that is varied systematically in our simulations. In our simulations, the linker regions participate in inter-linker interactions only. The strength of this inter-linker $\epsilon_{ns}$ in our simulations was varied in the range of typical strengths of short-range interactions in biomolecules. We vary the $\epsilon_{ns}$ for pair-wise inter-linker interaction (per pair) in the range of 0.1 kcal/mol ($\approx$0.2 kT) to 0.5 kcal/mol ($\approx$1 kT)~\cite{Sheu2003}. For comparison, strong non-covalent interactions such as the H-bonds are known to be of the order of 0.5 kcal/mol to 1.5 kcal/mol in solvated proteins. Similar values of isotropic, short-range interactions have been used in conventional coarse-grained protein force-fields such as the MARTINI~\cite{Marrink2007,Monticelli2008}.

\subsubsection*{LD Simulations Details.}
The dynamics of these coarse-grained polymers was simulated using the LAMMPS molecular dynamics package~\cite{Plimpton1995}. In these simulations, the simulator solves for the Newtons’s equations of motion in presence of a viscous drag (modeling effect of the solvent) and a Langevin thermostat (modeling random collisions by solvent particles)~\cite{Schneider1978}. The simulations were performed under the NVT ensemble, at a a temperature of 310 K. The mass of linker beads was set to 110 Da while the mass of the idealized functional domains (red and yellow beads in Fig.1) was set to 7000 Da that is approximately equal to the mass of the SH3 domain. The size of the linker beads was set at 4.2 A (of the same order as amino acids) while that of the functional domains was set at 20 A (~size of a folded SH3 domain~\cite{Musacchio1992}). The viscous drag was implemented via the damping coefficient, $\gamma = m/6\pi\eta a$. Here, m is the mass of an individual bead, `$\eta$’ is the dynamic viscosity of water and ‘a’ is the size of the bead. An integration time step of 15 fs was used in our simulations, and the viscosity of the surrounding medium was set at the viscosity of water. Similar values of these parameters have been previously employed for coarse-grained Langevin dynamics simulations of proteins~\cite{Bellesia2009}. 

\subsection*{Kinetic Monte carlo simulations}
To assess biologically relevant time-scales, we develop a phenomenological kinetic model wherein the individual multi-valent polymer chains are modelled as diffusing particles with fixed valencies. Each particle in our lattice Monte carlo simulations is a coarse-grained representation of the bead-spring polymer chains in the LD simulations, with an effective valency that is a simulation variable. Particles occupying adjacent sites on the lattice experience a weak, non-specific interaction (of strength $\epsilon_{ns}$) that is isotropic in nature. This isotropic attractive force is analogous to the inter-linker interactions in the LD simulations. A functional bond can stochastically form between any pair of neighbouring particles, provided both the particles possess unsatisfied valencies. The rate of bond formation between a pair of particles with unsatisfied valencies is $k_{bond}$. On the other hand, an existing functional bond can break with a rate $k_{break}$ that is equal to $k_{bond}$$exp-(\epsilon_{sp})$ in magnitude. $\epsilon_{sp}$  is the strength of each functional interaction. Assuming that there are unoccupied neighbouring sites, the monomers can diffuse on the lattice in any of the four directions (in 2D) with a rate $k_{diff}$. Additionally, the entire cluster that any given monomer is a part of can diffuse in either of the 4 directions with a scaled diffusion rate that is inversely proportional to the size of the cluster. A particle that is part of a cluster can diffuse away from the cluster with a rate $k_{diff} exp-(\sum\epsilon_{sp} + \epsilon_{ns}))$. Here, $\sum\epsilon_{sp} + \epsilon_{ns}$ is the magnitude of net interactions that any particle is involved in. These rates are schematically described in Fig.6.
Using these set of phenomenological rates, we allow the system to evolve using the exact stochastic simulation method or the Gillespie kinetic Monte Carlo algorithm~\cite{Gillespie1977}]. This algorithm has previously been used to model biological processes as diverse as gene regulation~\cite{Parmar2013} and cytoskeletal filament growth~\cite{Ranjith2009}. In this approach, a set of 'N' rates is initiated for any given current configuration (set of coordinates of each particle and their valencies). In other words,  10 potential events are initiated for each particle on the lattice. Namely, the 10 events per particle are a) 4 diffusion events, b) 4 cluster diffusion events, c) bond formation and d) bond breakage events. For any given configuration, all or only a subset of these events could be possible. We then advance the state of the system by executing one reaction at a time. The probability of each event is equal to its rate, $r_{i}^j$/$\sum \sum r_{i}^{j}$. Here i and j refer to the identity of the monomer and the event type, respectively. Given these set of propensities, we choose the event to be executed by drawing a uniformly distributed random number  and by comparing how the random number compares with the event propensities. The simulation time is advanced using the following expression, $\delta t$=$-(1/r_{total})ln(z_{2})$, where $z_{2}$ is a uniform random number. This $\delta t$ is based on the assumption that the waiting times between any two events are exponentially distributed. The above algorithm has to be iterated several times such that each reaction has been fired multiple times, suggesting the system has reached steady-state behaviour.
\subsubsection*{Rationale behind phase parameters in KMC simulations:}
Two key phase parameters in our KMC simulations are the valency per interacting particle and the bulk density of particles on the lattice. We performed simulations for valencies ranging from 3 to 6, typical of multivalent proteins enriched in membrane-less organelles. Unlike the non-specific which can only be one per neighbor, thereby a maximum of 4 per particle for a 2D-lattice, there could be more than one specific interactions between a pair of neighboring particles, as long as both the participating members have unsatisfied valencies. The bulk density of particles is defined as, $\phi$ = $N_{mono}$/$L^{2}$, where $N_{mono}$ and L are the number of particles and lattice size, respectively). In our kMC simulations, we varied $\phi$ within the range of 0.01 to 0.1, a range that is consistent with the analogous parameter for LD simulations, the occupied volume fractions within the LD simulation box ($\phi_{LD}$= $N_{mono}$.4/3$\pi$$R_{mono}^{3}$). For instance, a free monomer concentration ($C_{mono}$) of 10 uM corresponds to a $\phi_{LD}$ of 0.008 which increases to 0.17 for 200 uM. Further, the phase diagrams for KMC simulations were primarily computed for a weak non-specific $\epsilon_{ns}$ of 0.35 kT, chosen in order to focus on specific interaction-driven cluster growth. In the kMC simulations, $\epsilon_{ns}$ refers to the net non-specific interaction for a pair of monomers as opposed to a pair of interacting linker beads in the LD simulations. It must, however, be noted that the interaction strength for a pair of interacting chains is not merely additive (with the length of the polymer chain). A potential explanation for this could be found in earlier studies wherein it has been argued that smaller segments (~ length of 7-10 amino acids) within a long, solvated polymer chain (like IDPs) could behave like independent units referred to as blobs~\cite{Pappu2008}. Since the degree of coarse-graining employed in KMC is of the order of one particle per polymer chain, it lacks the microscopic degrees of freedom of the bead-spring polymer chain (in LD simulations). Hence we do not employ a higher non-specific interaction strength in our phase plot computations. As proof of principle, we show the mean pairwise interaction energies for dimers from the LD simulations, for a weak inter-linker interaction strength of 0.1 kcal/mol (Supplementary Fig.S7A) and a corresponding phase-diagram for non-specifically driven interactions (Supplementary Fig.S7B). All the mean cluster sizes in the MC simulations were computed over 300 independent kinetic Monte carlo trajectories. The mean molecular exchange times were computed over 1000-independent kMC runs.

\section{Acknowledgement}
This work was supported by NIH RO1 GM068670. We are grateful to Mark Miller (Durham University) and Ranjith Padinhateeri (Indian Institute of Technology Bombay) for useful discussions.

\bibliography{ref.bib}

\end{document}

\title{SUPPLEMENTARY INFORMATION: Liquid-liquid microphase separation leads to formation of membraneless organelles}

\author{Srivastav Ranganathan,$^\ast$ Eugene Shakhnovich $^\ast$}

% \address{$^\ast$ Department of Biosciences and Bioengineering, IIT Bombay, Powai, Mumbai 400076}

\maketitle

\markboth{Ranganathan et al.}{Physical determinants of condensate growth}

\begin{figure}
\centering
\captionsetup[subfigure]{labelformat=simple}
\includegraphics[scale=0.9]{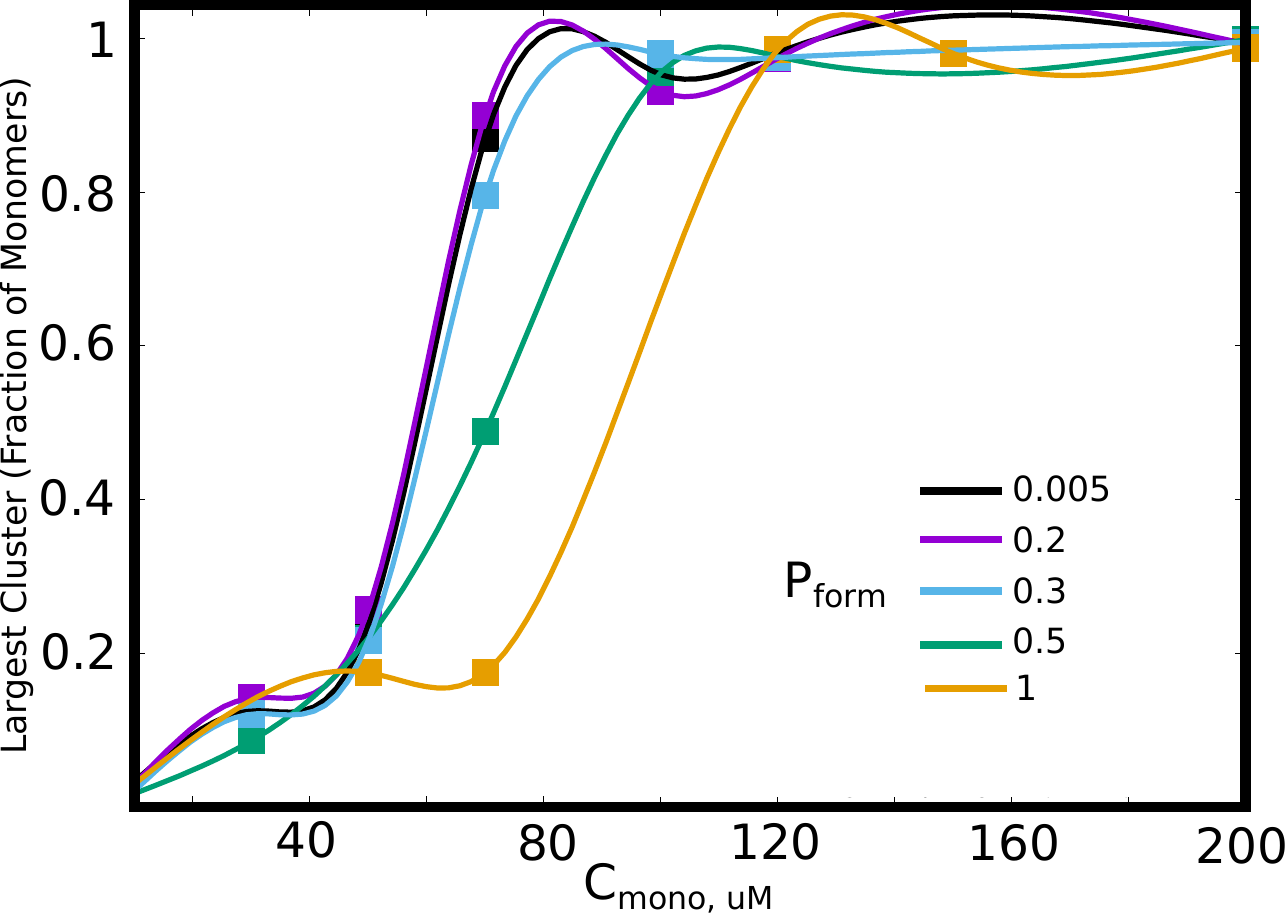}
\caption{The size of the largest cluster, for different values of bond formation probability, $P_{form}$. A lower $P_{form}$ results in a slower arrest of the clusters, and thereby results in increased cluster sizes for smaller free monomer concentration, $C_{mono}$.}
\label{fig:cluster_fraction_time_conc}
\end{figure}

\begin{figure}
\centering
\captionsetup[subfigure]{labelformat=simple}
\includegraphics[scale=0.43]{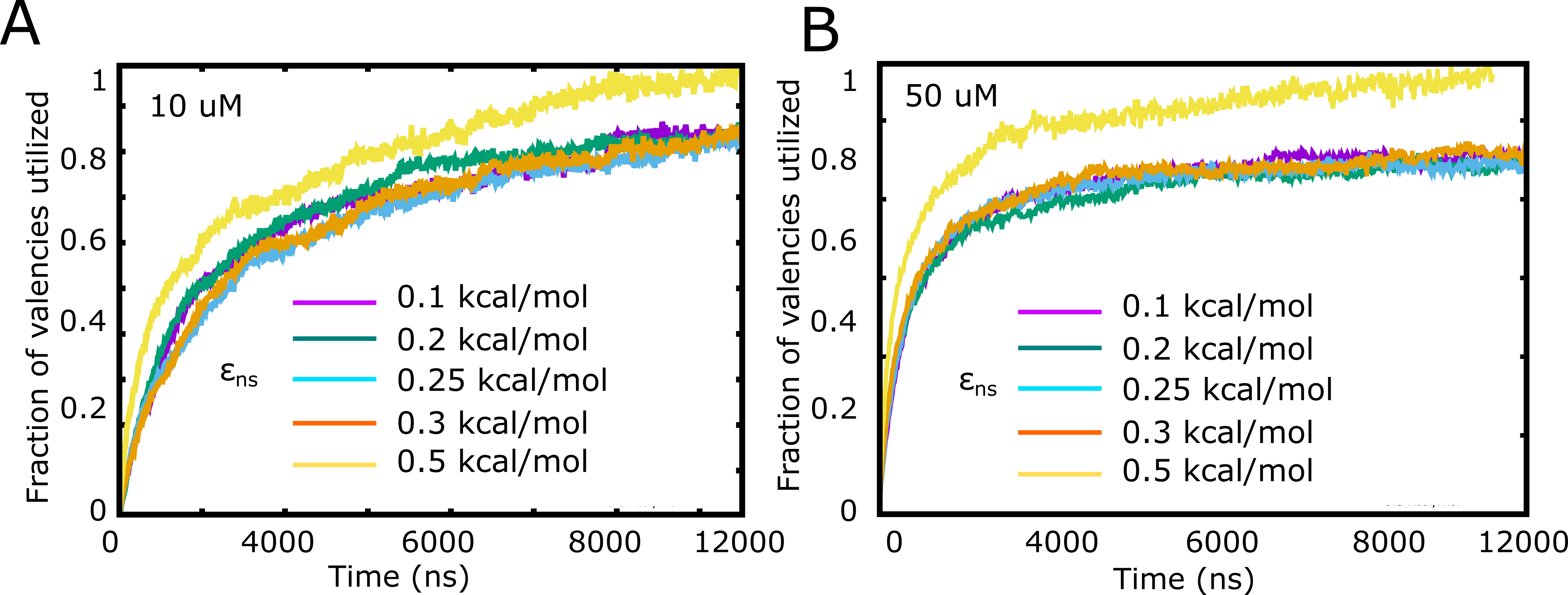}
\caption{Fraction of valencies utilized as a function of increasing inter-linker interaction strength, for A) 10 $\mu$M and B) 50 $\mu$M free monomer concentration.}
\label{fig:cluster_fraction_time_conc}
\end{figure}

\begin{figure}
\centering
 \captionsetup[subfigure]{labelformat=simple}
\includegraphics[scale=1]{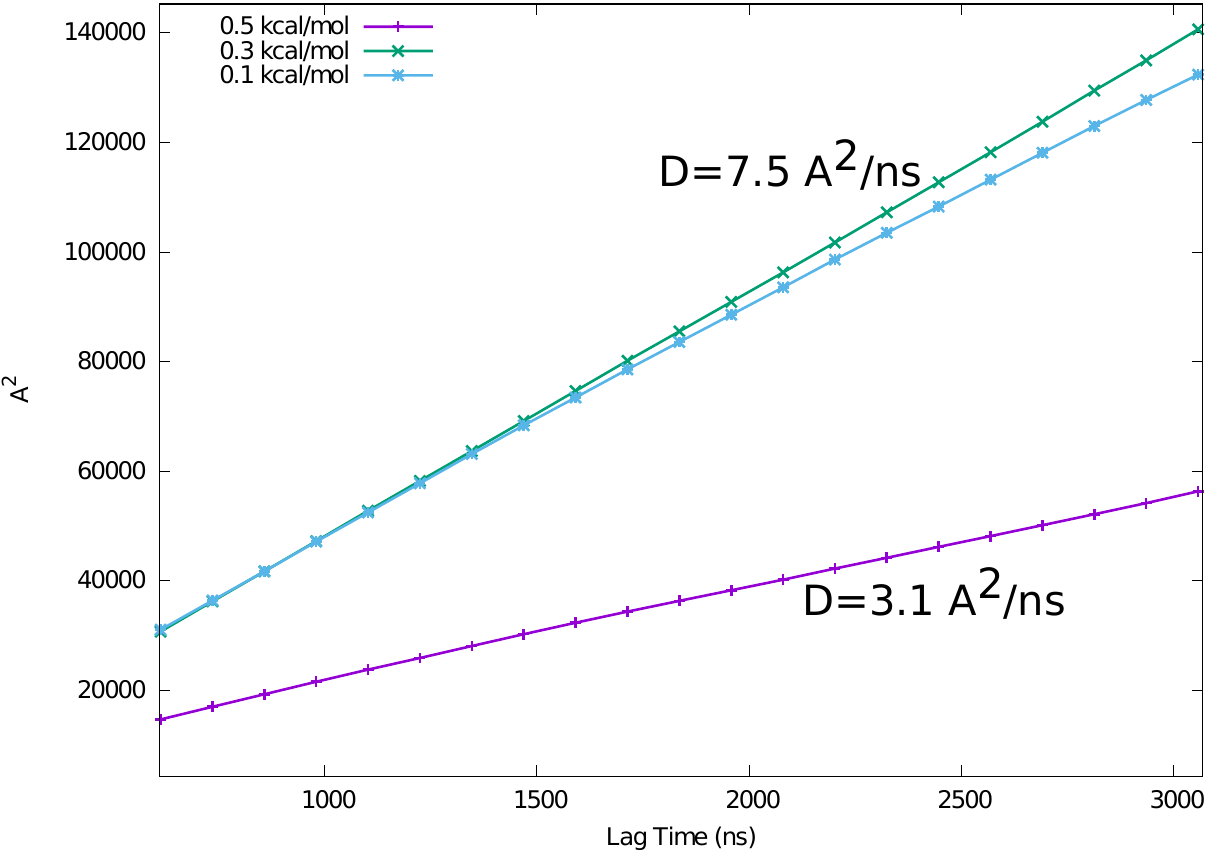}
\caption{The mean-squared displacements of the center of masses of the constituent polymer chains as a function of increasing lag-times, in the absence of functional interaction. The three curves show the lagtimes for varying values of isotropic interaction strength between the linker residues.}
\label{fig:control_msd}
\end{figure}

 \begin{figure}
 \centering
 \captionsetup[subfigure]{labelformat=simple}
 \subfloat[]{\includegraphics[scale=0.9]{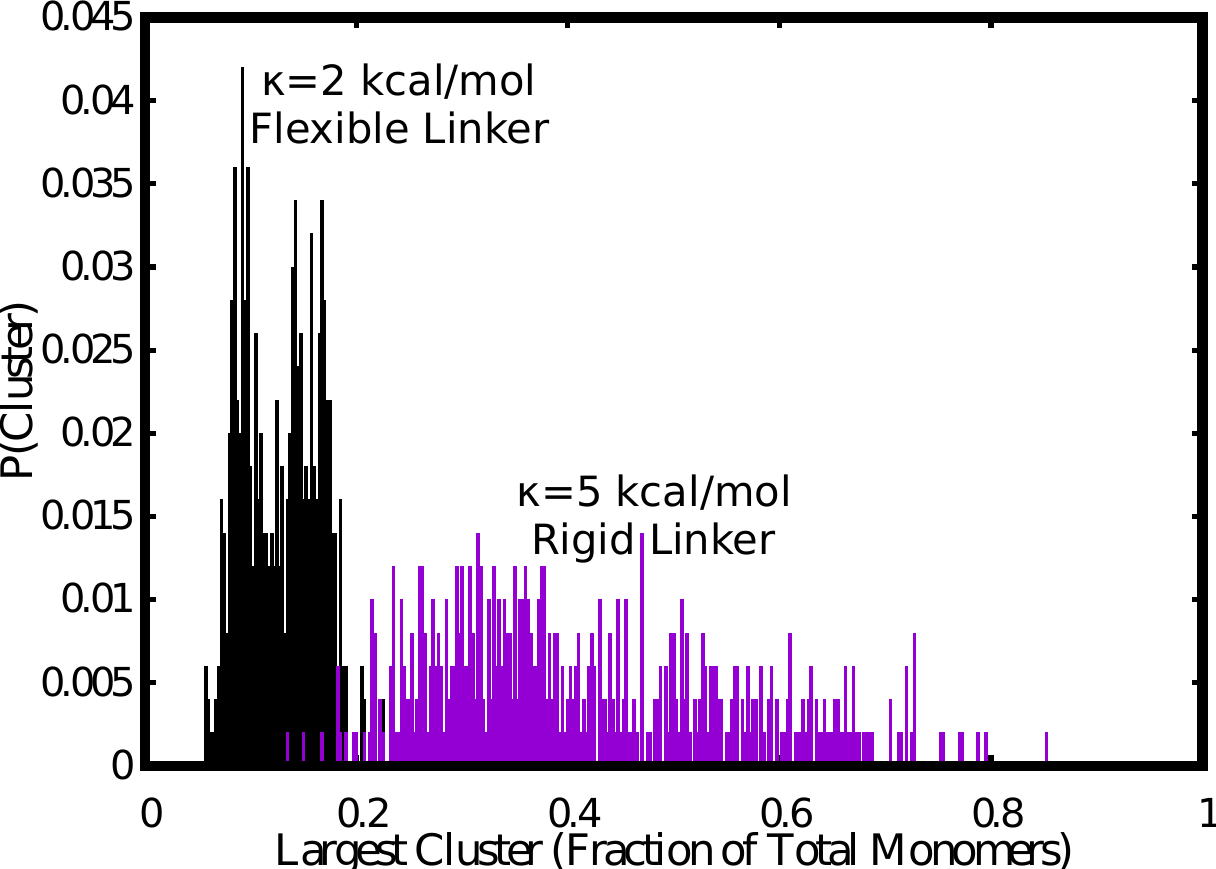}}
 \caption{Comparision between the size distributions of the largest cluster, for stiff ($\kappa$=2 kcal/mol) versus flexible ($\kappa$=5 kcal/mol) linker regions. The free monomer concentration used for this plot was 50 $\mu$M and an a weak interlinker interaction strength of $\epsilon_{n}$ = 0.1 kcal/mol was used.}
 \label{fig:aggre_no}
 \end{figure}

 \begin{figure}
 \centering
 \captionsetup[subfigure]{labelformat=simple}
 \subfloat[]{\includegraphics[scale=0.9]{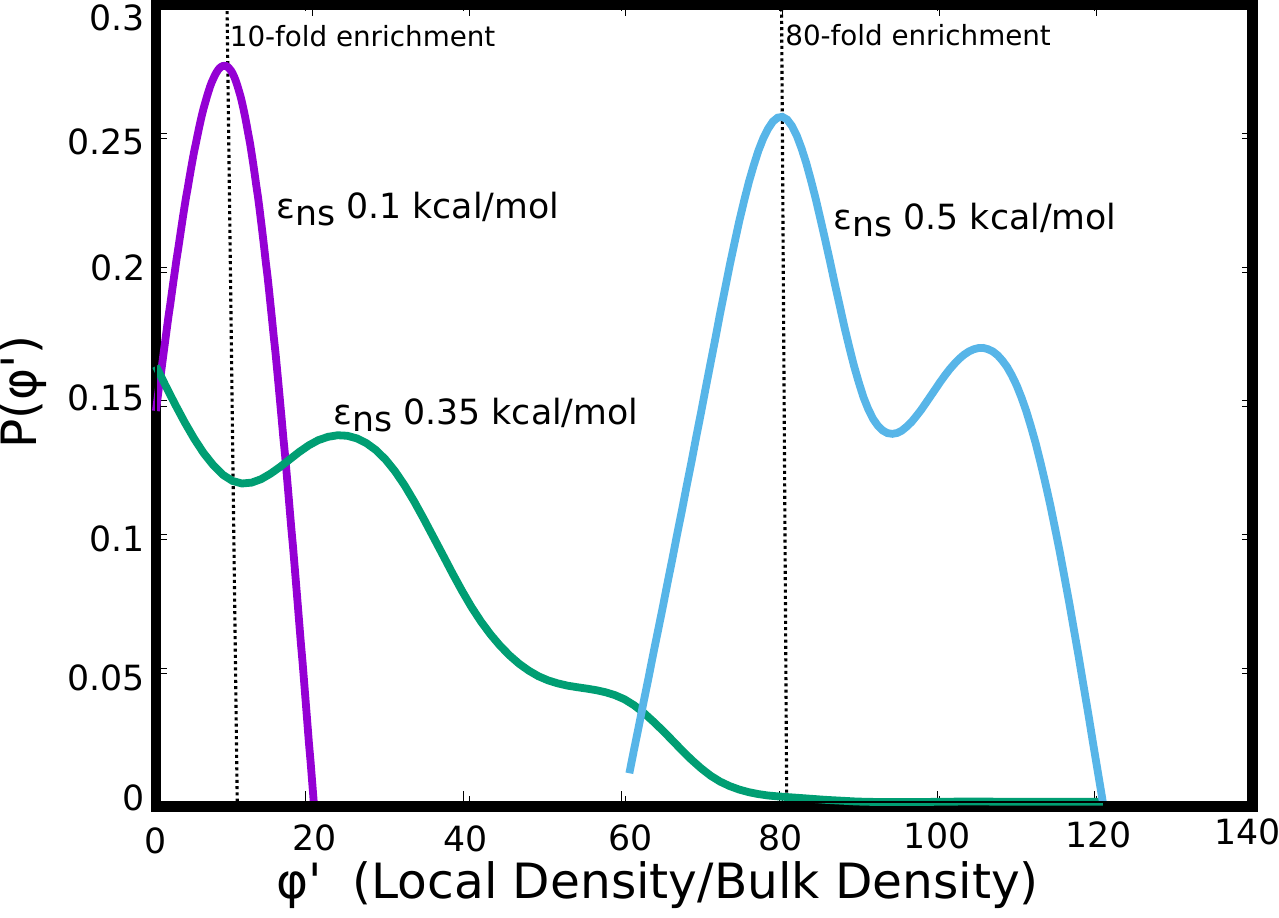}}
 \caption{The probability of finding clusters with varying densities (normalized by the bulk densities) for different values of inter-linker interactions. As the inter-linker interactions increase, the degree of enrichment can go from 10-fold to ~100-fold.}
 \label{fig:aggre_no}
 \end{figure}

 \begin{figure}
 \centering
 \captionsetup[subfigure]{labelformat=simple}
 \includegraphics[scale=0.6]{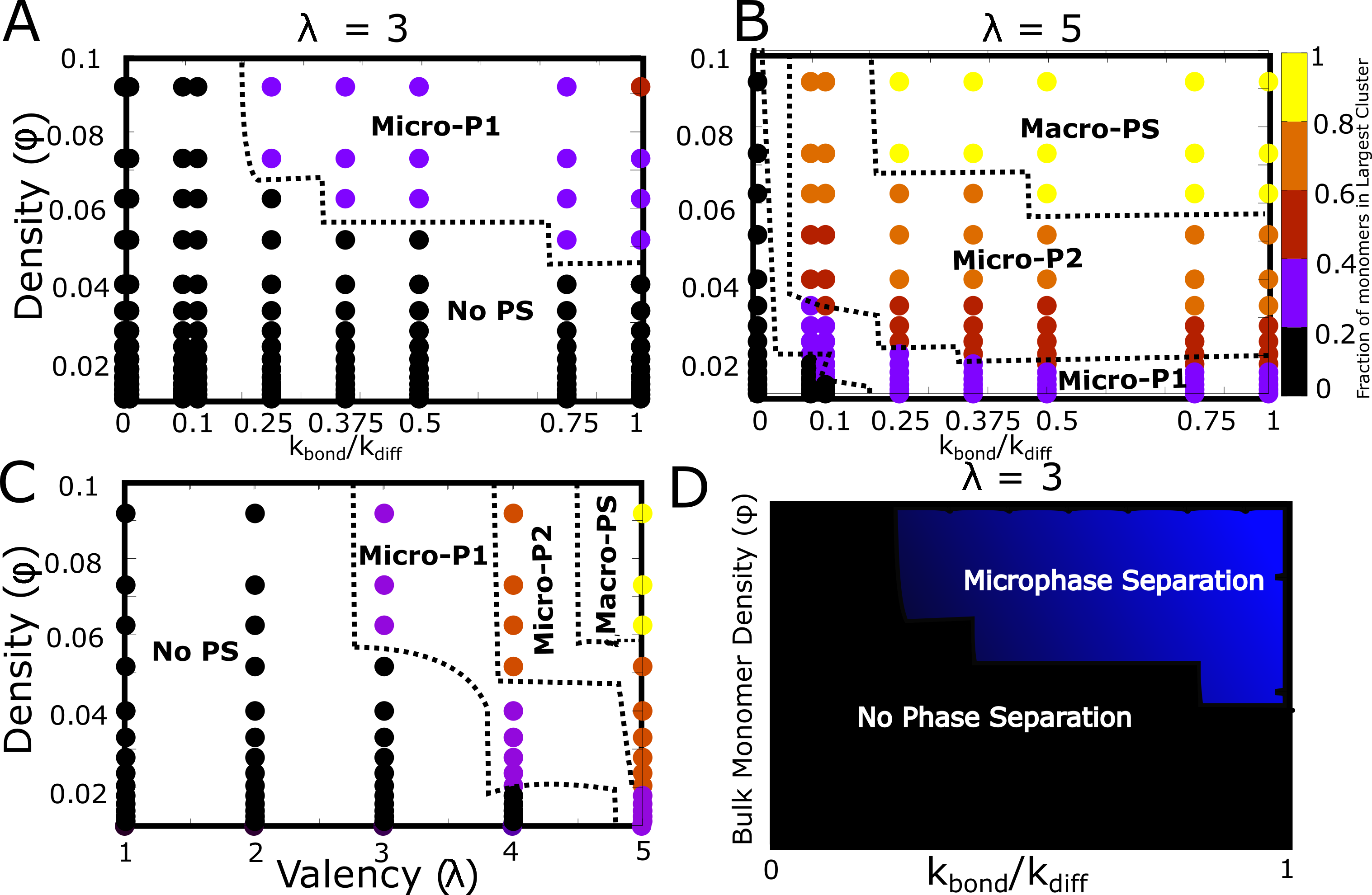}
 \caption{Detailed phase diagrams for A) and B) $\phi$-$k_bond$, C) $\phi$-$\lambda$ as the phase parameters. The cluster sizes were computed at the end of a simulation run of 2 hours (actual time), setting the rate of diffusion $k_diff$ to 1 $s^{-1}$. D) The bonding rate $k_{bond}$ was varied to identify the relationship between $k_{bond}$ and $k_diff$.}
 \label{fig:aggre_no}
 \end{figure}

 \begin{figure}
 \centering
 \captionsetup[subfigure]{labelformat=simple}
 \subfloat[]{\includegraphics[scale=0.7]{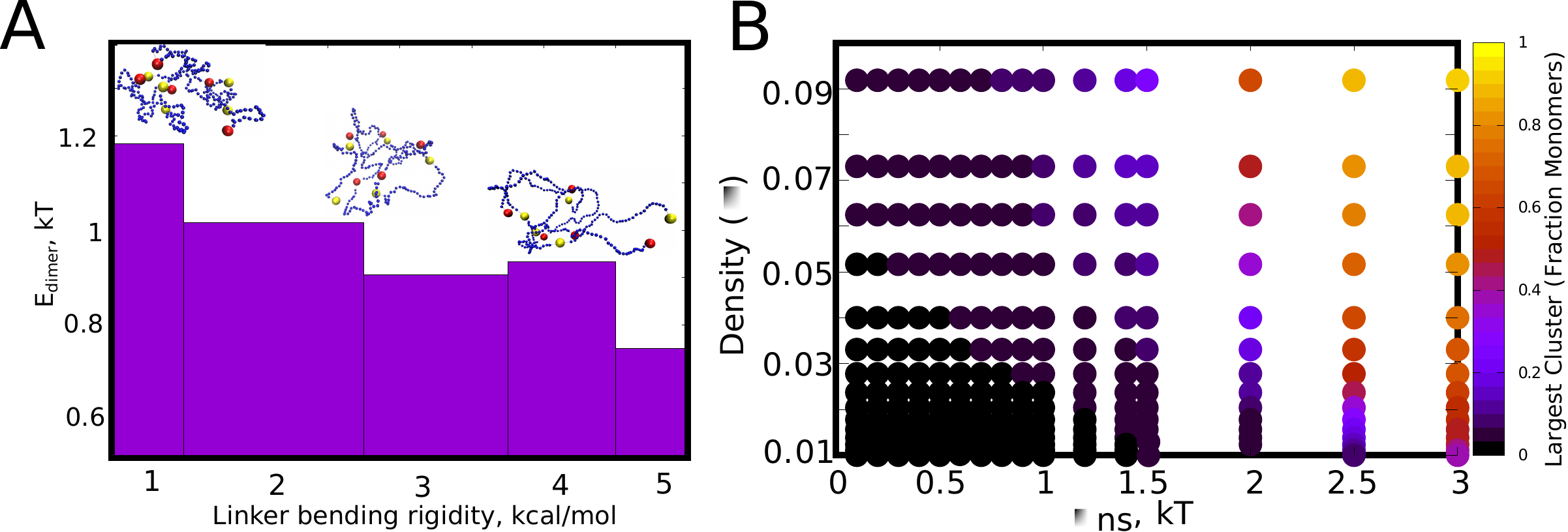}}
 \caption{A) The mean pair-wise interaction energy for 100 different dimeric structures (from the LD simulations), for an inter-linker interaction strength of 0.1 kcal/mol, for different values of linker bending rigidity. B) The $\epsilon_{ns}$-$\phi$ phase diagram (for $\lambda$=0) showing no phase separation for low values of isotropic interaction strength. However, for values of $\epsilon_{ns}$ $>$ 1kT, phase separation is observed at the end of the simulation timescale of 2 hours (actual time).}
 \label{fig:aggre_no}
 \end{figure}

 \begin{figure}
 \centering
 \captionsetup[subfigure]{labelformat=simple}
 \includegraphics[scale=0.6]{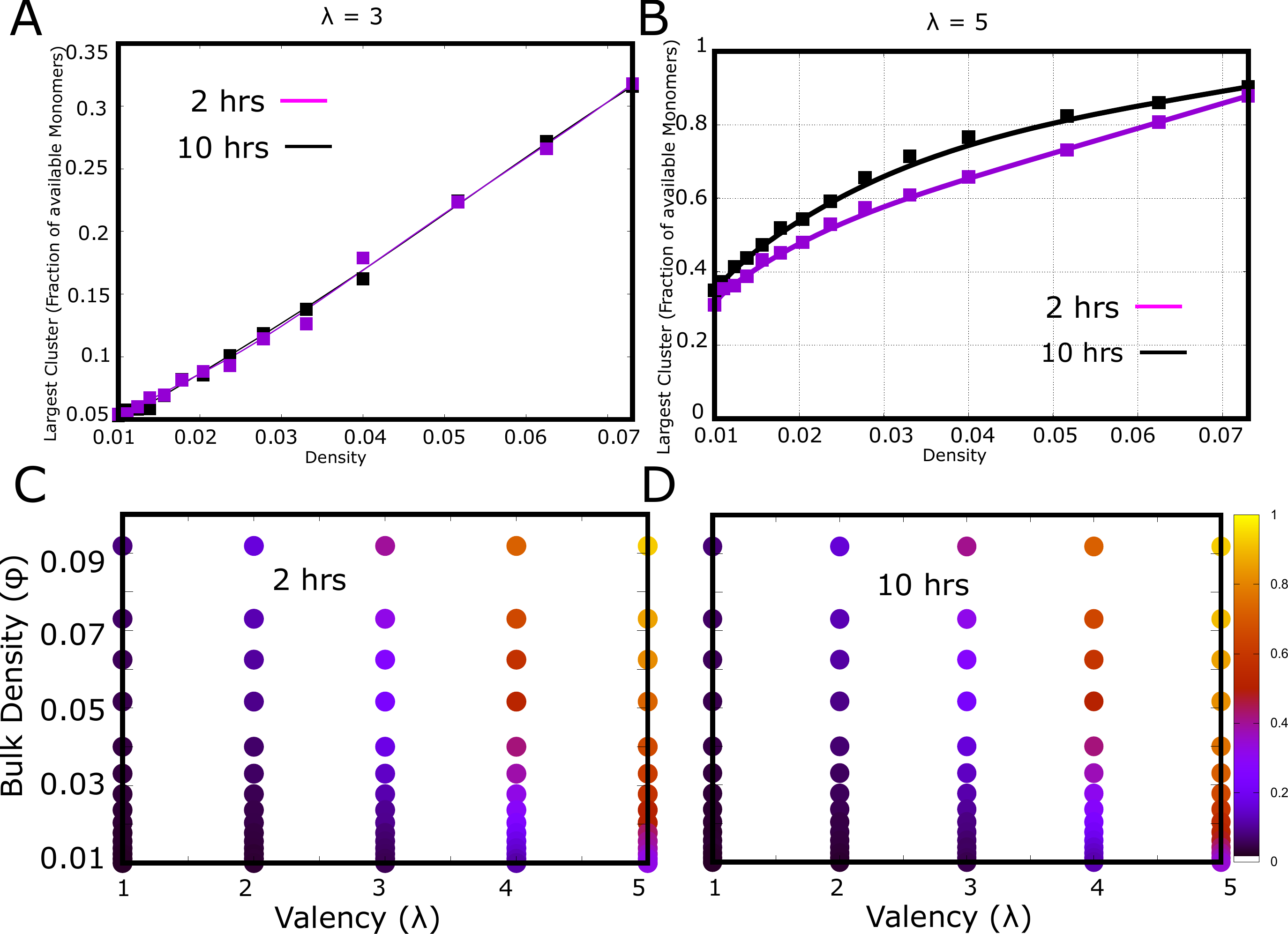}
 \caption{Convergence of phase diagrams. A) and B) shows the fraction of monomers in the largest cluster for 1 and 10 hours of actual time, for valency of 3 and 5, respecively. C) and D). The $\lambda$-$\phi$ phase diagram at the end of 2 and 10 hours of simulation time, respectively, showing very little difference.  }
 \label{fig:2h_v_10h}
 \end{figure} 

 \begin{figure}
 \centering
 \captionsetup[subfigure]{labelformat=simple}
 \includegraphics[scale=0.7]{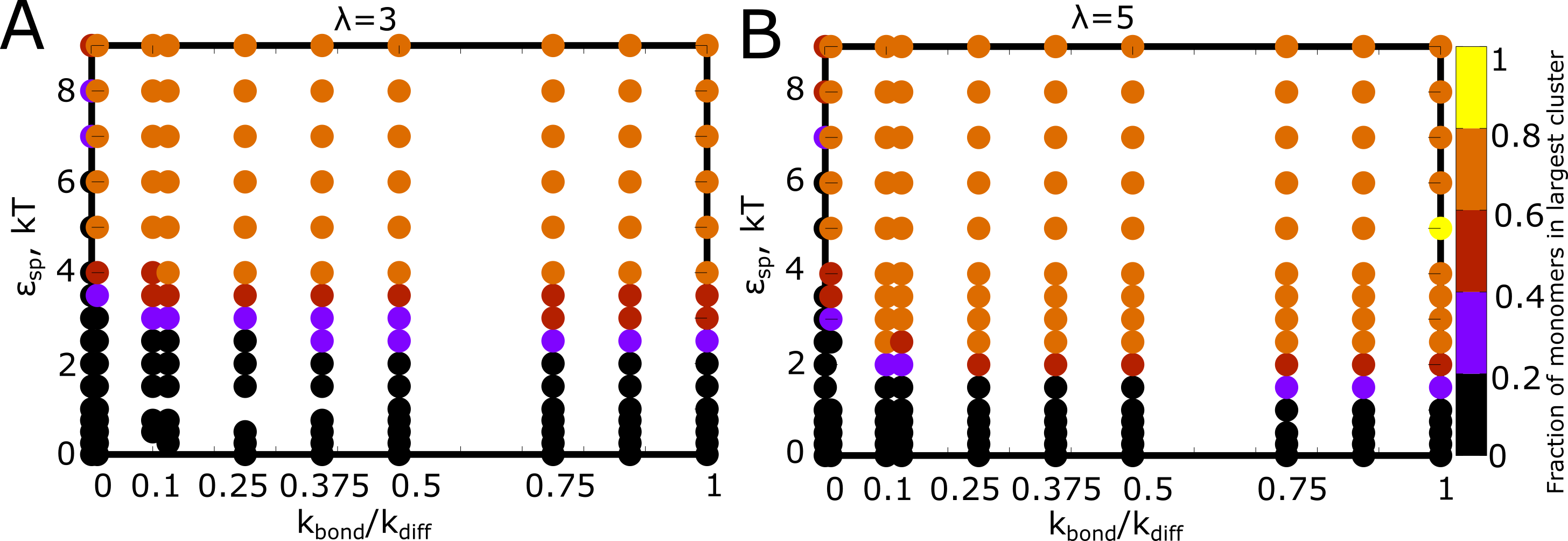}
 \caption{Detailed phase diagrams for A) and B) $\epsilon_{sp}$-$k_{bond}$ as the phase parameters. The bulk density of particles was set to 0.04, an intermediate density identified from the previous phase diagrams with density as a phase parameter. The cluster sizes were computed at the end of a simulation run of 2 hours (actual time), setting the rate of diffusion $k_diff$ to 1 $s^{-1}$. }
 \label{fig:kb_v_eps}
 \end{figure}

 \begin{figure}
 \centering
 \captionsetup[subfigure]{labelformat=simple}
 \includegraphics[scale=0.65]{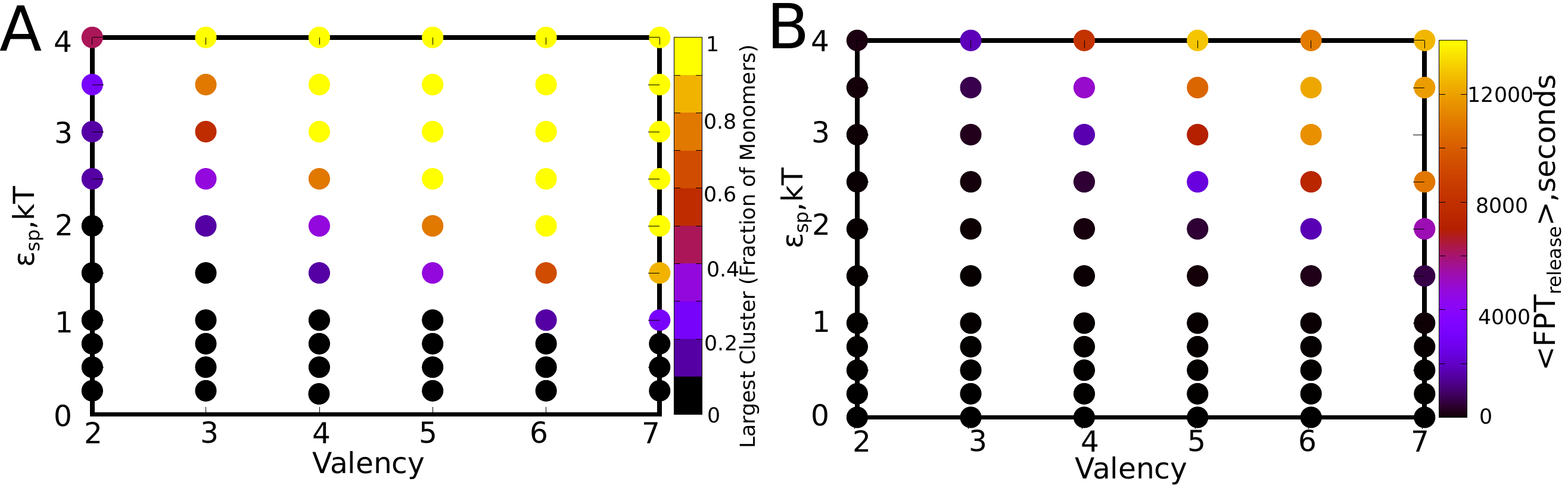}
 \caption{Mean cluster sizes for variation in $\epsilon_{sp}$ and $\lambda$, for a bulk density of 0.04, and a $k_{bond}$/$k_{diff}$ ratio of 1. B) Mean first passage times for a particle to exchange between a cluster and the bulk. The parameter values are same as in panel A.}
 \label{fig:aggre_no}
 \end{figure}

 \begin{figure}
 \centering
 \captionsetup[subfigure]{labelformat=simple}
 \includegraphics[scale=0.6]{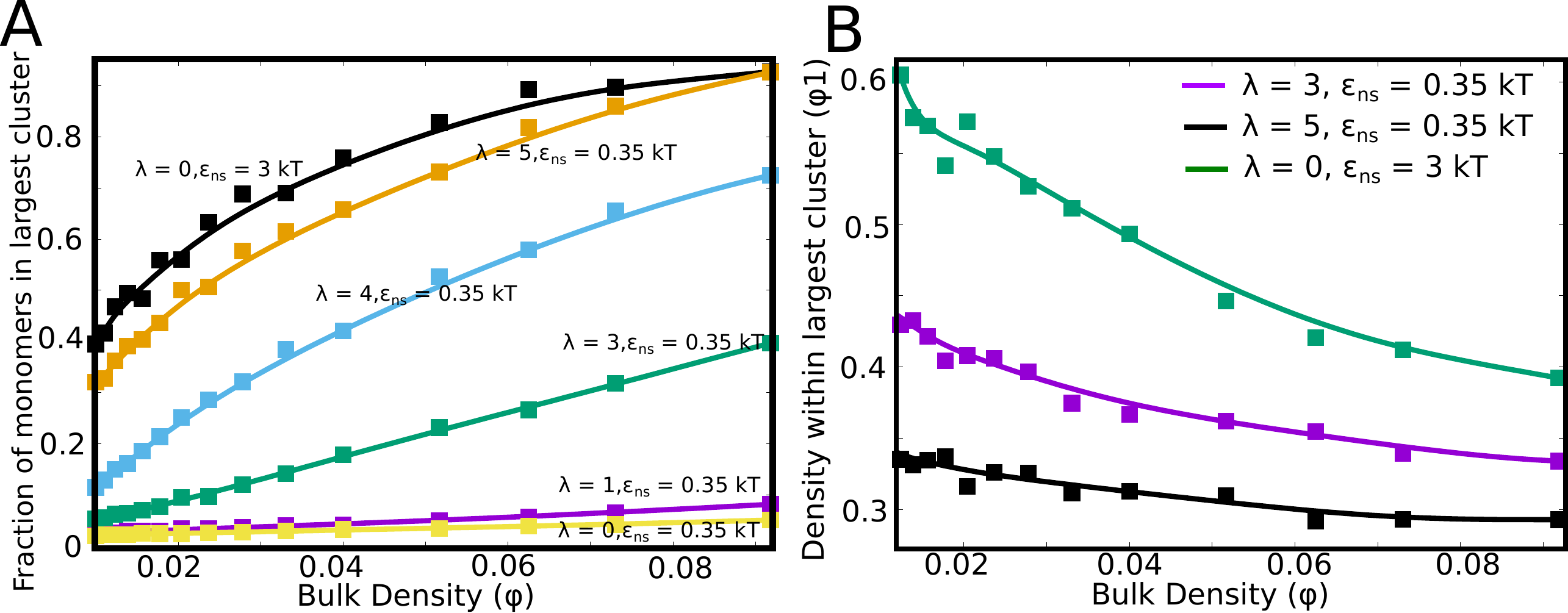}
 \caption{Cluster formation driven by isotropic versus specific interactions. A) Comparison of cluster sizes (as a fraction of total monomers) for different scenarios. The black curve shows cluster sizes for assembly driven by strong non-specific interactions alone ($\lambda$=0 and $\epsilon_{ns}$=3 kT). For a scenario involving weak isotropic interactions ($\epsilon_{ns}$ = 0.35 kT), the curves approach that of the isotropic interactions for higher valencies ($\lambda$ $\rightarrow$ 4). B) Densities of largest cluster for assemblies stabilized by isotropic interactions ($\epsilon_{ns}$ = 3kT, $\lambda$=0;green curve) and two different valencies (purple and black curves). Clusters stabilized by isotropic interactions are denser than the ones held together by specific interactions.}
 \label{fig:aggre_no}
 \end{figure}

% \begin{figure}
% \centering
% \captionsetup[subfigure]{labelformat=simple}
% \subfloat[]{\includegraphics[scale=0.57]{MC_Figures/CvKb_v3_mean.eps}}
% \hspace{0.2 cm}
% % \subfloat[]{\includegraphics[scale=0.65]{MC_Figures/CvKb_v5_mean.eps}}
% \subfloat[]{\includegraphics[scale=0.57]{MC_Figures/CvKb_v5_mean.eps}}
% \caption{Mean cluster sizes as a function of $k_{bond}^{+}$ and particle density ($N/L^{2}$) at the end of 1 hour are plotted in the form of a heat map (color codes represent mean cluster sizes as a fraction of total monomers). Sub-panels (A) and (B) are phase plots for particle valency of 3 and 5, respectively.}
% \label{fig:MC_phase_diag_Cvkb}
% \end{figure}
 
%\bibliography{resprop.bib}

%\section*{Acknowledgements}
%\section*{Author contributions statement}
%\section*{Additional Information}
%\section*{SUPPLEMENTARY MATERIAL}